\documentclass[12pt]{article}

\usepackage{graphicx}

\font\blackboard=msbm10 at 12pt
\font\blackboards=msbm7
\font\blackboardss=msbm5
\newfam\black
\textfont\black=\blackboard
\scriptfont\black=\blackboards
\scriptscriptfont\black=\blackboardss

\newcommand{\junk}[1]{}

\newcommand{\ba}{\begin{array}}
\newcommand{\ea}{\end{array}}
\newcommand{\be}{\begin{equation}}
\newcommand{\ee}{\end{equation}}
\newcommand{\bea}{\begin{eqnarray}}
\newcommand{\eea}{\end{eqnarray}}
\newcommand{\beas}{\begin{eqnarray*}}
\newcommand{\eeas}{\end{eqnarray*}}

\def\laplace{{\kern1pt\vbox{\hrule height 1.2pt\hbox{\vrule width
1.2pt\hskip
  3pt\vbox{\vskip 6pt}\hskip 3pt\vrule width 0.6pt}\hrule height
  0.6pt}
  \kern1pt}}
\def\scriptlap{{\kern1pt\vbox{\hrule height 0.8pt\hbox{\vrule width
  0.8pt
  \hskip2pt\vbox{\vskip 4pt}\hskip 2pt\vrule width 0.4pt}\hrule height
  0.4pt}
  \kern1pt}}

\def\roughly#1{\raise.3ex\hbox{$#1$\kern-.75em\lower1ex\hbox{$\sim$}}}

\newcommand{\gone}[1]{}

\begin{document}
\pagestyle{plain}
\setcounter{page}{1}

\baselineskip16pt

\begin{titlepage}

\begin{flushright}
ACT-07-12, MIFPA-12-15
\end{flushright}

\bigskip

\vspace{8 mm}

\begin{center}

{\Large \bf Momentum Modes of $M5$-branes in a $2d$ Space \\}

\end{center}

\vspace{7 mm}

\begin{center}

{\bf Shan Hu $^{1,a}$, and Dimitri Nanopoulos $^{1,2,3,b}$}

\vspace{3mm}
{\small \sl $^{1}$George P. and Cynthia W.Mitchell Institute for Fundamental Physics, Texas A\&M University,} \\
{\small \sl  College Station, TX 77843, USA} \\
{\small \sl $^{2}$Astroparticle physics Group, Houston Advanced Research Center (HARC),} \\

{\small \sl Mitchell Campus, Woodlands, TX 77381, USA} \\

{\small \sl $^{3}$Academy of Athens, Division of Nature Sciences,} \\

{\small \sl 28 panepistimiou Avenue, Athens 10679, Greece} \\

\vspace{3mm}

{\small \tt $^{a}$hushan@physics.tamu.edu, $^{b}$dimitri@physics.tamu.edu} \\

\end{center}

\vspace{8 mm}
\begin{abstract}

We study $M5$ branes by considering the selfdual strings parallel to a plane. With the internal oscillation frozen, each selfdual string gives a $5d$ SYM field. All selfdual strings together give a $6d$ field with $5$ scalars, $3$ gauge degrees of freedom and $8$ fermionic degrees of freedom in adjoint representation of $U(N)$. Selfdual strings with the same orientation have the SYM-type interaction. For selfdual strings with the different orientations, which could also be taken as the unparallel momentum modes of the $6d$ field on that plane or the $(p,q)$ $(r,s)$ strings on $D3$ with $(p,q)\neq (r,s)$, the $[i,j]+[j,k]\rightarrow [i,k]$ relation is not valid, so the coupling cannot be written in terms of the standard $N \times N$ matrix multiplication. 3-string junction, which is the bound state of the unparallel $[i,j]$ $[j,k]$ selfdual strings, may play a role here. 

\end{abstract}

\vspace{1cm}

Keywords: Field Theories in Higher Dimensions, Brane Dynamics in Gauge Theories, M-Theory

\begin{flushleft}

\end{flushleft}
\end{titlepage}
\newpage

\section{Introduction}

The effective theory on $M5$ branes is special in that the basic excitations, the selfdual strings, are $1d$ objects other than the $0d$ objects, like those on $D$ branes or $M2$ branes \cite{01,02}. If the selfdual strings can be closed and can shrink to point like the fundamental strings, then we will still have a theory with the semi-point-like excitations. For selfdual strings without the charge, this is indeed the case. In abelian theory, the quantization of the point-like $M2$ confined to $M5$ brane gives the $(2,0)$ tensor multiplet \cite{11,12,13}. Moreover, the basic excitations on $(2,0)$ little string theory \cite{2} living on $N$ coincident type IIA $NS5$ branes are closed fundamental strings, which are also the closed selfdual strings coming from $M2$ wrapping the M theory circle intersecting $NS5$ along a closed curve.

There is no evidence showing that for selfdual strings carrying charge, the situation is the same. Consider $D4$ branes, which are $M5$ branes compactified on $x_{5}$. The $[i,j]$ monopole string on $D4$ can carry the $D0$ charge. The closed $[i,j]$ monopole string with the vanishing length carrying $D0$ will appear as the point-like instanton with charge $[i,j]$. However, the classical instanton solution with charge $[i,j]$ is associated with the $[i,j]$ monopole string extending along a straight line, while the point-like\footnote{By point-like, we mean the instanton solution is localized in $R^{4}$, centered around a point.} $1/2$ BPS instanton solutions are always chargeless. The closed monopole string with no $D0$ charge is the selfdual string with the winding number and the momentum both zero along $x_{5}$, which will give a $5d$ massless SYM field in adjoint representation of $U(N)$, in addition to the original $5d$ SYM field coming from the selfdual strings winding $x_{5}$ once. The SYM field like this is always massless even if the $D4$ branes are separated from each other, so it will dominate at the Coulomb branch. However, on $D4$, no such field exists. We do not get the clue for the existence of the closed charged selfdual strings. Actually, when the charged selfdual string becomes curved, different parts of it may exert force to each other, so it cannot vibrate freely and cannot be closed as the chargeless strings do.

Then we have to incorporate the $1d$ object in a $6d$ field theory. In this paper, we will take the $[i,j]$ tensionless selfdual string extending along, for example, the $x_{5}$ direction, as the point-like $6d$ excitation, which is in the position eigenstate in $1234$ space but in the $P_{5}=0$ momentum eigenstate in $x_{5}$. If so, selfdual strings extending along the same direction cannot give the complete Hilbert space for the 6d particle. To get the full Hilbert space, we need to consider selfdual strings with the orientations covering all directions in a plane. The superposition of the selfdual strings parallel to a plane can give the 6d point-like excitations localized in 12345 space, but it seems that somehow, the position representation is not the suitable one, since it is the $[i,j]$ selfdual strings other than the $[i,j]$ point-like excitations that naturally exist. One plane is already enough to define a $6d$ field theory, so different planes may give the U-dual versions of the same $6d$ theory. This is quite similar with the ${\cal N}=4$ SYM theory, for which, one $(p,q)$ string defines a $4d$ field theory, while the rest $(p,q)$ strings give the S-dual $4d$ theories.

Theories with the line-like excitations are intrinsically different from those with the point-like excitations. If the excitations are line-like, a reduction on $x_{5}$ will give the selfdual strings extending along $x_{5}$, while a further reduction on $x_{4}$ will make $P_{4}=0$. The first reduction selects a particular selfdual string; the second one is just the ordinary reduction in local field theories. With $4$ and $5$ switched, we will get the selfdual strings extending along $x_{4}$ with $P_{5}=0$, which is S-dual to the selfdual strings extending along $x_{5}$ with $P_{4}=0$. On the other hand, if the excitations are point-like, both sequences will give the point-like selfdual strings with $P_{4}=P_{5}=0$. In \cite{WI1, WI2}, Witten has shown that due to the conformal symmetry, the $45$ and $54$ reductions of the $6d$ $(2,0)$ theory will give two S-dual $4d$ SYM theories other than one $4d$ theory, which strongly indicates that the basic excitations on $M5$ cannot be point-like.

Recall that in \cite{3}, the equations of motion for the 3-algebra valued $(2,0)$ tensor multiplet contain a constant vector field $C_{\mu}$. A given $C_{\mu}$ will reduce the dynamics from $6d$ to $5d$. However, if $C_{\mu}$ covers all directions in a plane, we will get a set of $\theta$-parameterized $5d$ SYM theories\footnote{A single $5d$ SYM theory already contains the complete KK modes of the $6d$ theory \cite{41,42}. The KK modes are realized as the field configurations in $R^{4}$ carrying the nonzero Pontryagin number. Here, the $\theta$-parameterized SYM field must have the zero Pontryagin number, because it is the zero mode of the $6d$ field along the $C_{\mu}(\theta)$ direction.}, which is equivalent to a $6d$ theory with $5$ scalars, $3$ gauge degrees of freedom and $8$ fermionic degrees of freedom. Each $5d$ SYM theory has the $5d$ vector multiplet in adjoint representation of $U(N)$, arising from the quantization of the open $[i,j]$ $M2$ intersecting $M5$ along the $C_{\mu}(\theta)$ direction. The oscillation along $C_{\mu}(\theta)$ is frozen, so the spectrum is the same as that from the quantization of the open string. For the given $6d$ $(2,0)$ tensor multiplet field configuration, each $5d$ SYM field comes from the reduction of the $6d$ field along $C_{\mu}(\theta)$. This is actually a special kind of KK compactification, using the polar coordinate other than the rectangular coordinate.

Suppose the selfdual string orientations are restricted in $45$ plane, then the common eigenstates of $[\hat{X}_{1},\hat{X}_{2},\hat{X}_{3},\hat{P}_{4},\hat{P}_{5}]$ could be selected as the bases to generate the Hilbert space. In this respect, it is convenient to consider the KK mode of the $6d$ $(2,0)$ theory with $x_{4}$ and $x_{5}$ compactified to circles with the radii $R_{4}$ and $R_{5}$. $M5$ branes with the longitudinal $x_{4}$ and $x_{5}$ compactified is dual to the $D3$ branes with the transverse $x'^{45}$ compactified. The vacuum expectation values of the 2-form field on $x_{4} \times x_{5}$ is converted to the transverse positions of the $D3$ branes on $x'^{45}$. The duality differs from the T-duality in that two longitudinal dimensions are converted to one transverse dimension so the total dimensions are reduced from $11$ to $10$. The $(n/R_{4},m/R_{5})$ momentum mode of the $6d$ theory is dual to the $(p,q)$ string winding $x'^{45}$ $k$ times, with $n=kp$, $m=-kq$, $p$ and $q$ co-prime. The $[i,j]$ $(p,q)$ strings form the adjoint representation of $U(N)$. Correspondingly, the momentum modes as well as the original $6d$ field are also in the $U(N)$ adjoint representation.

The $(n/R_{4},m/R_{5})$ momentum modes are in $4d$ vector multiplet $V_{4}$, which, when combine together, give the $6d$ tensor multiplet $T_{6}$ in $U(N)$ adjoint representation. Although the $6d$ tensor multiplet is in the adjoint representation, the coupling involving more than two fields cannot be realized as the standard matrix multiplication. SYM coupling is obtained by studying the scattering amplitude of the open strings ending on $D$ branes. $[i,j]+ [j,k] \rightarrow [i,k]$, so the $L^{i}_{j}M^{j}_{k}N^{k}_{i}$ type coupling is possible. On the other hand, for $M5$, we need to consider the scattering of the selfdual strings parallel to a given plane. Selfdual strings with the same orientation still have the SYM-coupling, while for selfdual strings with different orientations, the $[i,j]+[j,k] \rightarrow [i,k]$ relation is not valid, so the $[i,j]$ $(n/R_{4},m/R_{5})$ mode, the $[j,k]$ $(k/R_{4},l/R_{5})$ mode, and the $[k,i]$ $(-(n+k)/R_{4},-(m+l)/R_{5})$ mode of the $6d$ field cannot couple unless $nl=mk$, in which case all of them belong to the same $5d$ SYM theory with $\tan \theta = -\frac{nR_{5}}{mR_{4}}$. The difference between the SYM theory and the effective theory on $M5$'s is rooted in the fact that the boundary of the open string is the point, while the boundary of the open $M2$ is the line.

The $(n/R_{4},m/R_{5})$ momentum mode of the $6d$ SYM theory on $D5$ is dual to the open $F1$ ending on $D3$ with the winding number $(n,m)$ around $x'_{4} \times x'_{5}$. From the scattering amplitude of the massive winding open strings and the massless open strings on $D3$, one can reconstruct the original $6d$ SYM theory. Similarly, for $M5$, we need to consider the interaction of the open $(p,q)$ strings ending on $D3$ winding $x'^{45}$ $k$ times for all co-prime $(p,q)$ and all nonnegative $k$, or in other words, the interaction for all of the monopoles and dyons in ${\cal N}=4$ SYM theory. When the scalar fields on $M5$ branes get the vacuum expectation value, $D3$ branes will be separated in the rest $5d$ transverse space, while the 3-string junctions \cite{51,52}, which are also the bound states of the $[i,j]$ $[j,k]$ selfdual strings each carrying the transverse momentum in $45$ plane, can be formed. The 3-string junction is characterized by three vectors $(r_{4},r_{5})$, $(s_{4},s_{5})$ and $(t_{4},t_{5})$ in $45$ plane, which may couple with the $(n/R_{4},m/R_{5})$ momentum modes as long as $(n,m)\propto (r_{4},-r_{5})$, or $(n,m)\propto (s_{4},-s_{5})$ or $(n,m)\propto (t_{4},-t_{5})$. The quantization of the 3-string junction with the lowest spin content gives the $1/4$ BPS multiplet $V_{4} \otimes ([1/2]\oplus[0]\oplus[0])$ with $2^{6}$ states \cite{6}, which, when lifted to $6d$, becomes the $(2,1)$ multiplet with $2^{7}$ states, among which, half are tri-fundamental and half are tri-anti-fundamental representation of $U(N)$. $[i,l]+[l,j,k]\rightarrow [i,j,k]$, $[j,m]+[i,m,k]\rightarrow [i,j,k]$, $[k,n]+[i,j,n]\rightarrow [i,j,k]$, so we may have couplings like $V^{i}_{l}T_{ijk}T^{ljk}$ or $V^{i}_{l}T_{ijk}T^{ljn} V^{k}_{n}$. \footnote{In \cite{1111}, the scattering amplitude involving two charged $5d$ KK modes and one $5d$ zero mode for $M5$ branes compactified on $S^{1}$ is discussed. The charged KK mode is the $1/4$ BPS dyonic instanton in $5d$ massive $(2,1)$ multiplet, while the zero mode is in $5d$ massless vector multiplet. The incorporation of the spin-3/2 particles of the $(2,1)$ multiplet into the theory requires a novel fermionic symmetry.}

However, with the given scalar vacuum expectation value $\vec{v}_{i}$ on $M5$, the bound states of the $[i,j]$ $\vec{v}_{ij}$ $(n/R_{4},m/R_{5})$ momentum mode and the $[j,k]$ $\vec{v}_{jk}$ $(k/R_{4},l/R_{5})$ momentum mode exist only under the certain condition $h(\vec{v}_{ij},\vec{v}_{jk},n/R_{4},m/R_{5},k/R_{4},l/R_{5}  )>0$ with $h=0$ specifying the marginal stability curve \cite{7}. Especially, when $\vec{v}_{ij} = 0$, $\forall \: i,j$, except for $n=m=0$ or $k=l=0$, which is at the curve of the marginal stability, the rest bound states do not exist. The bound state of the $[i,j]$ $(0,0)$ momentum mode and the $[j,k]$ $(n/R_{4},m/R_{5})$ momentum mode could be taken as the tensionless selfdual string carrying the longitudinal momentum. It is unclear whether it is the bound state or just two separate states.

When compactified on $x_{5}$, the $[i,j]$ selfdual string extending along $x_{5}$ becomes the $[i,j]$ $F1$ localized in $1234$ space. For the rest selfdual strings, to get the definite $P_{5}$ momentum, they must carry the definite transverse momentum thus are projected into the bound states of the $[i,j]$ $F1$ and the $[i,j]$ monopole string, each carrying the suitable $P_{4}$ and $P_{5}$ transverse momentum, localized in $123$ space. The $1/2$ BPS solutions for the BPS equations in $5d$ SYM theory match well with the above states, except for instantons, which is also localized in $1234$ space. Similarly, the 3-string junctions in $6d$, when projected to $5d$, become the bound states of the $[i,j]$ $F1$ and the $[j,k]$ monopole string with $P_{4}$ and $P_{5}$ transverse momentum, localized in $123$ space. Except for the dyonic instantons, the generic $1/4$ BPS solutions in $5d$ SYM theory involving no more than three $D4$'s have a one-to-one correspondence with these states.

The rest of this paper is organised as follows: In section 2, we discuss the longitudinal momentum mode on branes with special emphasis on the point-like charged $1/2$ BPS instantons on $D4$ branes. In section 3, we consider the selfdual strings on $M5$ branes parallel to the $45$ plane, or equivalently, the KK momentum mode of the $6d$ theory upon the compactification on $x_{4} \times x_{5}$. In section 4, we study the interaction of the $6d$ $(2,0)$ theory by considering its KK mode on $x_{4} \times x_{5}$. In section 5, we consider various momentum-carrying BPS states in $5d$ SYM theory, especially, the monopole string carrying the longitudinal momentum and the string carrying $D0$ charge. In section 6, we discuss the states living at the triple intersection of $M5$ branes. The discussion is in section 7.

\section{The point-like charged $P_{5}$ momentum mode on $D4$ and the nonabelian $5d$ massive $(2,0)$ tensor multiplet}

In this section, we discuss the longitudinal momentum mode on $M5$ branes, which is carried by selfdual strings. We will argue that there is no closed charged selfdual strings and so no classical point-like charged momentum mode on $M5$ branes. Nevertheless, on $N$ coincident $D4$ branes, the localized $[i,j]$ $P_{5}$ mode (D0 brane) could be realized as the quantum superposition of the $[i,j]$ $P_{5}$ mode in momentum eigenstate of, for example, $P_{4}$, which is the $[i,j]$ tensionless monopole string wrapping $x_{4}$, carrying the $D0$ charge as well as the $P_{4}$ longitudinal momentum, or equivalently, the $[i,j]$ $(n,1)$ string on $D3$ if $x_{4}$ is compactified. The $5d$ massive nonabelian $(2,0)$ tensor multiplet with mass $P_{5} = 1/R_{5}$ could then be decomposed into a tower of $4d$ massive $U(N)$ vector multiplet arising from the quantization of the $(n,1)$ open strings for all $n \in \textbf{Z}$.

\subsection{The longitudinal momentum mode on branes}

Let us first see the transverse momentum of the branes. For a $Dp$ brane with the transverse dimension $x_{p+1}$ compactified to a circle $S^{1}$, the brane may locate at a particular point in $S^{1}$ or have the definite momentum along $S^{1}$. In the former situation, after the T-duality transformation along $x_{p+1}$, $Dp$ becomes $D(p+1)$ with the gauge field $A_{p+1}$ getting the vacuum expectation value. In the latter case, the T-duality transformation converts $Dp$ into $D(p+1)$ with the definite electric flux $F_{0(p+1)}$. $D(p+1)$ cannot have the definite $A_{p+1}$ and $F_{0(p+1)}$ simultaneously, just as $Dp$ cannot have the definite $X_{p+1}$ and $P_{p+1}$ at the same time. In M theory, if $x_{5}$ is compactified to $S^{1}$, $M2$ transverse to $x_{5}$ may either locate at a particular point in $S^{1}$ or have the definite $P_{5}$ momentum. $M2$ with zero $P_{5}$ momentum is $D2$ in type IIA string theory. $M2$ with nonzero $P_{5}$ momentum is the $D2$-$D0$ bound state. The transverse velocity should be the same everywhere on $D2$, so the $D0$ charges are uniformly distributed over $D2$. If the masses of the $D2$ and $D0$ are $m_{2}$ and $m_{0}$ respectively, the energy of the $D2$-$D0$ bound state is $\sqrt{m_{2}^{2}+m_{0}^{2}}$, in contrast to the energy of the $D4$-$D0$ bound state, which is $m_{4}+m_{0}$.

As for the longitudinal momentum mode on branes, consider $Dp$ brane with the longitudinal dimension $x_{p}$ compactified to a circle $S^{1}$, $Dp$ may carry momentum along $x_{p}$. Under the T-duality transformation in $p$ direction, we get the $D_{p-1}$-$F1$ bound state with $F1$ ending on $D_{p-1}$ winding the transverse circle $x'^{p}$. If the $i_{th}$ and the $j_{th}$ $D_{p-1}$ branes are separated along another transverse dimension $x'^{p+1}$, the closed $F1$ becomes open, which, under the T-duality transformation along $x'^{p}$, gives the open $[i,j]$ string ending on $Dp$ branes carrying the $P_{p}$ momentum. So, more precisely, the $P_{p}$ longitudinal momentum of the $Dp$ brane is the $P_{p}$ transverse momentum of the open strings living on them. Consider the $i_{th}$ and the $j_{th}$ $Dp$ branes with the $[i,j]$ string orthogonally connecting them carrying momentum $P_{p}$, the total energy is $m_{p}+ \sqrt{T^{2}_{F1}|\vec{v}_{i}-\vec{v}_{j}|^{2}+P_{p}^{2}}$, where $m_{p}$ and $T_{F1}|\vec{v}_{i}-\vec{v}_{j}|$ are masses of $Dp$ and $[i,j]$ string respectively. When $\vec{v}_{i}=\vec{v}_{j}$, the energy reduces to $m_{p}+ |P_{p}|$, corresponding to the $Dp$ branes carrying the $[i,j]$ $P_{p}$ longitudinal momentum. The low energy effective action on $N$ coincident $Dp$ branes is the $U(N)$ SYM theory. When compactified on $x_{p}$, one may get an infinite tower of KK modes still in the adjoint representation of $U(N)$. We have seen that the $P_{p}$ momentum carries charge, thus is indeed in the adjoint representation.

For $M2$ branes, two $M2$ branes orthogonally intersecting at a point may form the threshold bound state, so the transverse momentum of one $M2$ gives the longitudinal momentum of the other, as long as the two can keep intersecting at one point. Similarly, it is natural to expect that for $M5$ branes, the $P_{k}$ longitudinal momentum may actually be the $P_{k}$ transverse momentum of the selfdual strings living in them. Especially, for $5d$ SYM theory, the $P_{5}$ momentum may just come from the monopole strings living in $D4$. However, $P_{5}$ like this is distributed in a straight line other than localized at a point. Localized instantons do exist, which are $D0$ branes resolved in $D4$. The line-like $P_{5}$ momentum carried by selfdual strings has the $[i,j]$ charge, while the point-like $P_{5}$ momentum corresponding to $D0$ branes is chargeless. It is necessary to construct the point-like $1/2$ BPS momentum mode carrying charge.

One may want to consider the closed selfdual strings, which, with the length shrinking to zero, may carry the point-like momentum. However, the selfdual strings carrying charge must extend along a straight line. We cannot get the closed charged selfdual strings unless the worldvolume of the $M5$ branes has the nontrivial 1-cycle. Consider the $[i,j]$ selfdual string segment extending along $ABC$, where $A$ $B$ $C$ are three points in $M5$. Suppose $AB \bot BC$, then the $AB$, $BC$ strings are actually the same as the $[i,j]$ $F1$ and the $[i,j]$ $D1$. The configuration like this is not BPS, so $F1$ and $D1$ may exert force to each other. The $[i,j]$ $F1$-$D1$ bound state is not at the threshold and has the mass $|\vec{v}_{ij}|\sqrt{|AB|^{2}+|BC|^{2}}$ due to the binding energy. We are actually talking about the $[i,j]$ selfdual string segment extending along $AC$. The $[i,j]$ selfdual string cannot vibrate freely, because different parts may exert force to each other.

On the other hand, the chargeless selfdual strings do not have this problem. They can be closed and may carry the point-like momentum. One such example is the $[i,i]$ selfdual string. On the $i_{th}$ $M5$ brane, we have the zero length $[i,i]$ closed selfdual string, or in other words, the collapsed $M2$ brane, the quantization of which gives the expected $U(1)$ $(2,0)$ tensor multiplet \cite{11,12,13}. When compactified on $x_{5}$, the point-like $D2$ with $P_{5}$ momentum becomes the $D0$ confined to the $i_{th}$ $D4$. Another example is the little string theory \cite{2, 81,82}. Consider $N$ coincident type IIA $NS5$ branes with the longitudinal dimension $x_{5}$ compactified to a circle with the radius $R_{5}$. The $11_{th}$ dimension is $x_{10}$ which is compactified with the radius $R_{10}$. There are closed fundamental strings with tension $2 \pi R_{10} T_{M2}$ living in $NS5$, which are closed $M2$'s wrapping $x_{10}$ intersecting $M5$ along a closed curve. After a series of duality transformations, the momentum mode (carried by the type IIA string) along $x_{5}$ is converted to the $D0$ branes living in $D4$ branes with the compactified transverse dimension $x'_{5}$. If the original type IIA closed string has the finite length, we will get a closed $D2$ wrapping $x'_{5}$ intersecting $D4$ along a closed curve carrying the uniformly distributed $D0$ charge. The $D0$ branes are obtained when the size of the $D2$ brane carrying them shrinks to zero. Actually, the type IIA $NS5$ brane picture and the $D4$ brane picture are S-dual to each other with $5$ and $10$ switched. On type IIA $NS5$ branes, the $P_{5}$ momentum is carried by the closed string. When the string shrinks to a point, we simply take it as a momentum mode without the string involved. Similarly, on $D4$ branes, we may have closed $D2$ carrying $P_{5}$ momentum. With the closed $D2$ shrinking to a point, we are left with the $D0$ brane/$P_{5}$ momentum.

On a single $NS5$ brane, purely $P_{5}$ momentum is carried by strings that do not wind $x_{5}$, or alternatively, $M2$'s that do not wrap $x_{5}$. Correspondingly, on a single $D4$ brane, the purely $P_{5}$ momentum mode should be carried by the closed $D2$ branes other than strings. The complete KK modes on $NS5$ branes upon the compactification on $x_{5}$ are characterized by $(m,n)$, where $m$ and $n$ are the winding number and the momentum mode of the string along $x_{5}$ respectively. The $(m,n)$ mode has the mass $2 \pi m T_{F1}R_{5}+n/R_{5}$. $(m,0)$ mode, $(0,n)$ mode and $(m,n)$ mode are in the $5d$ $(1,1)$, $(2,0)$ and $(2,1)$ multiplets preserving $1/2$, $1/2$, and $1/4$ supersymmetries respectively \cite{9}. For the dyonic strings in \cite{42}, with $x_{6}$ compactified to a circle, the bound state of the $[1,2]$ and $[2,1]$ dyonic strings carrying $n$ instanton number is just be the $(1,n)$ mode here. The generic $(m,n)$ mode is obtained by the quantization of the type IIA strings. Especially, with $N_{L}=0$, the level-matching condition requires $N_{R}=mn$, so the oscillation mode along the string must be turned on \cite{9,10}. This is easy to understand. For type IIA string wrapping $x_{5}$, $P_{5}$ can only come from the internal oscillation since there is no transverse momentum along $x_{5}$. On the other hand, if the $[i,j]$ selfdual strings wrapping $x_{5}$ cannot oscillate, the $P_{5}$ momentum carried by it can only come from selfdual strings extending in $1234$ space. The $(m,n)$ mode in this case is actually the threshold bound state of the $(m,0)$ mode and the $(0,n)$ mode. The former is associated with the selfdual string extending along $x_{5}$, while the latter is given by the tensionless selfdual strings extending along $1234$ space carrying the $P_{5}$ momentum. It is possible for the $[i,j]$ $(m,0)$ mode and the $[j,k]$ $(0,n)$ mode to form the threshold bound state with $ijk$ indices, which we will discuss later.

\subsection{The point-like charged $P_{5}$ momentum mode}

Now, let us consider the relation between the point-like charged $P_{5}$ momentum mode and the line-like charged $P_{5}$ momentum mode. For $N$ coincident $D4$ branes with the longitudinal dimension $x_{4}$ compactified, the T-duality transformation along $x_{4}$ converts the $D4$-$D0$ bound state into the $D3$-$D1$ bound state with $D1$ winding $x'^{4}$. More precisely, $D1$ carrying the definite electric flux, or equivalently, $D1$-$F1$ bound state, corresponds to $D0$ in $P_{4}$ momentum eigenstate, while $D1$ with the definite $A_{4}$ field corresponds to $D0$ in $X_{4}$ position eigenstate. We may take the $(n,1)$ strings in $D3$ as the bases, the superposition of which gives $D1$ with the definite $A_{4}$, which is also the $D0$ located at a definite point in $D4$. Moreover, since $D1$ ending on $D3$'s can also carry charge, the $[i,j]$ $D1$ wrapping $x'^{4}$ is dual to the $[i,j]$ $D2$ wrapping $x_{4}$ with the $D0$ charge spreading over $\vec{v}_{ij} \times x_{4}$. When $\vec{v}_{ij}=0$, we are left with the tensionless $[i,j]$ monopole string winding $x_{4}$ carrying the uniformly distributed $D0$ charge, which could also be taken as the $[i,j]$ $D0$ in $P_{4}$ momentum eigenstate with the eigenvalue $0$. Similarly, the $[i,j]$ $(n,1)$ string is dual to the combination of the tensionless $[i,j]$ monopole string winding $x_{4}$ carrying the uniformly distributed $D0$ charge and the massless $[i,j]$ string carrying $P_{4}$ momentum, which could be simply taken as the $[i,j]$ $D0$ with the nonzero $P_{4}$ eigenvalue. Still, the superposition of the $[i,j]$ $(n,1)$ strings for all $n$ gives the $[i,j]$ $D0$ in $X_{4}$ position eigenstate. The instanton solutions describing the $[i,j]$ $D0$ with the definite $P_{4}$ momentum have the translation invariance along $x_{4}$, involving both magnetic and the electric fields. These states compose the complete spectrum for the charged $D0$ living in $D4$, while the localized charged $D0$ is the superposition of them.

The above conclusion can be stated in the language of $M5$ branes, since the bound state of the tensionless $[i,j]$ monopole string wrapping $x_{4}$ carrying $D0$ charge and the massless $[i,j]$ string carrying the $P_{4}$ momentum is just the tensionless selfdual string living in $x_{4} \times x_{5}$ carrying the transverse $(P_{4},P_{5})$ momentum. Consider $M5$ branes with $x_{4}$ and $x_{5}$ compactified with the radii $R_{4}$ and $R_{5}$, $B_{45} = \frac{1}{2 \pi R_{4}R_{5}}$. Tensionless selfdual string winding $x_{4}$ and $x_{5}$ $m$ and $n$ times may carry the transverse momentum $(-n/R_{4}, m/R_{5})$, thus could be described by the wave function 
\begin{equation}\label{0111}
	\frac{1}{4\pi^{2} R_{4}R_{5}}\exp \left\{i(-\frac{nx_{4}}{R_{4}}+\frac{mx_{5}}{R_{5}})\right\} \delta(x_{1}-X_{1})\delta(x_{2}-X_{2})\delta(x_{3}-X_{3}). 
\end{equation}
The $P_{5}$ momentum localized in $x_{4}$ has the wave function 
\begin{equation}
	\frac{1}{2\pi R_{5}} \exp \left\{\frac{i mx_{5}}{R_{5}}\right\} \delta(x_{1}-X_{1}) \delta(x_{2}-X_{2})\delta(x_{3}-X_{3})\delta(x_{4}-X_{4}), 
\end{equation}
which is the superposition of (\ref{0111}) with all $n\in \textbf{Z}$. In this respect, at least for $M5$ with at least two dimensions compactified, the longitudinal momentum is still given by the basic excitations, which are the selfdual strings here. The only difference is that the selfdual string is the one dimensional object, so the transverse momentum carried by it will appear as the two dimensional wave other than the one dimensional wave like the momentum carried by particles. To get the complete spectrum, we need the particles with the location covering $x_{4}$, or the selfdual strings with the orientation covering the $45$ plane. $\left\{\delta(x_{4}-X_{4})| X_{4}\in [0, 2\pi R_{4}) \right\}$ and $ \left\{\frac{1}{2\pi R_{4}}e^{\frac{i nx_{4}}{R_{4}}} | n\in \textbf{Z} \right\}$ are different bases for the same Hilbert space.

On $D4$ branes, the $[i,j]$ $P_{5}$ momentum can only be carried by the $[i,j]$ selfdual strings. There is no classical solution for the point-like instanton with the charge $[i,j]$. However, we do have the solution for the chargeless point-like instantons, which may consist of $N$ instanton partons with charge $[1,2]$, $\cdots$, $[N-1,N]$, $[N,1]$, while the size $\rho$ is the parameter characterizing the distance between the instanton partons \cite{111,112}. Similarly, for type IIA $NS5$ branes with $x_{5}$ compactified, the $P_{5}$ momentum is carried by the point-like closed strings, which are also composed by the $[1,2]$, $\cdots$, $[N-1,N]$, $[N,1]$ closed selfdual strings from M theory's point of view. It is difficult to get a single closed selfdual string with charge $[i,j]$. However, if one longitudinal dimension of $D4$ is compactified, an instanton on $D4$ branes will be dual to a D-string on $D3$ branes winding the transverse circle one time. A closed D-string is composed by the $[1,2]$, $\cdots$, $[N-1,N]$, $[N,1]$ D-string segments \cite{111,112}. The $[i,j]$ D-string segment can exist independently, because it is the $[i,j]$ monopole string extending along the compactified longitudinal dimension carrying the transverse $P_{5}$ momentum. Similarly, for the $P_{4}$ longitudinal momentum mode on $D4$ branes, we have the $[i,j]$ $P_{4}$ mode carried by the $[i,j]$ open string, which is the $[i,j]$ selfdual string winding $x_{5}$ one time, carrying the transverse $P_{4}$ momentum thus could also exist separately. The $[1,2]$, $\cdots$, $[N-1,N]$, $[N,1]$ $P_{4}$ modes can combine together to give a chargeless $P_{4}$ mode as well, but it is not necessary anymore.

\subsection{The $5d$ $U(N)$ massive $(2,0)$ tensor multiplet}

For $N$ coincident $M5$ branes with the compactified $x_{5}$, the $6d$ $(2,0)$ tensor multiplet could be decomposed into the zero mode and the KK modes. The zero mode is in $5d$ $U(N)$ vector multiplet, while the KK modes are in $5d$ massive $(2,0)$ tensor multiplets. The point-like $[i,j]$ $P_{5}$ mode (the $[i,j]$ $D0$ brane) is the the superposition of the tensionless $[i,j]$ selfdual strings extending in, for example, the $45$ plane, carrying the transverse momentum $(P_{4},P_{5})$ with the same $P_{5}$ but all $P_{4}$. Corresponding, the $5d$ $(2,0)$ tensor multiplet is then decomposed into the sum of the $4d$ KK modes in $U(N)$ vector multiplet, arising from the quantization of the above selfdual string states.

Consider $B_{\mu\nu}$ in $5d$ $U(1)$ massive $(2,0)$ multiplet with mass $1/R_{5}$. $B_{\mu\nu}$ satisfies the selfduality condition
\begin{equation}\label{sel}
	B_{\mu\nu} = -\frac{iR_{5}}{2}\epsilon_{\mu\nu\lambda\rho\sigma}\partial^{\lambda}B^{\rho\sigma}
\end{equation}
and the equation of motion
\begin{equation}
	\partial^{\lambda}\partial_{\lambda}B_{\mu\nu} + \frac{1}{R^{2}_{5}}B_{\mu\nu}=0,
\end{equation}
where $\mu,\nu,\lambda,\rho,\sigma = 0, 1,2,3,4$ \cite{42}. Do a further compactification on $x_{4}$,
\begin{equation}
	B_{\mu\nu} = \sum_{k}e^{ikx_{4}/R_{4}}B^{(k)}_{\mu\nu}. 
\end{equation}
Due to (\ref{sel}), for $i,j = 0,1,2,3$ and $k \in Z$, $B^{(k)}_{ij}$ could be expressed in terms of $B^{(k)}_{i4}$ thus could be dropped. We are left with a tower of the $4d$ massive vector field $A^{(k)}_{i} = B^{(k)}_{i4}$ satisfying the constraint 
\begin{equation}\label{zx}
\partial^{i}A^{(k)}_{i} = 0
\end{equation}
as well as the equation of motion
\begin{equation}\label{zy}
\partial_{j}\partial^{j}A^{(k)}_{i}+(\frac{1}{R^{2}_{5}}+\frac{k^{^{2}}}{R^{2}_{4}})A^{(k)}_{i}=0. 
\end{equation}
Each $A^{(k)}_{i}$ carries $3$ degrees of freedom, the same as $B^{(k)}_{ij}$. In $45$ plane, the $(n,1)$ string carries the momentum $(n/R_{4}, 1/R_{5})$ thus gives the $4d$ vector multiplet $A^{(n)}_{i}$. All of the $A^{(n)}_{i}$ are on the equal footing, which is consistent with the S-duality. To account for the $P_{5}$ momentum $m/R_{5}$ with $m>1$, we need $(n,m)$ strings, so altogether, all $(n,m)$ strings should be included to give the complete $6d$ dynamics. Under the compactification on $x_{4}$ and $x_{5}$, the $6d$ field $B_{\alpha \beta}$ with $\alpha, \beta = 0,\cdots,5$ is decomposed into the $4d$ KK modes $(n/R_{4}, m/R_{5})$ corresponding to the $(n,m)$ string. Each KK mode gives a $4d$ massive vector field $A^{(n,m)}_{i}$, for which, the constraint and the equation of motion could be obtained by replacing $R_{5}$ and $k$ in (\ref{zy}) by $R_{5}/m$ and $n$.

Extending the discussion to the nonabelian case is a little difficult, since we don't know the equations for the nonabelian tensor field. However, we do know that the $(n/R_{4},0)$ mode, which is the KK mode of the $5d$ massless SYM field, is in $4d$ adjoint massive vector multiplet. The rest $(n/R_{4},m/R_{5})$ modes are related with $(n/R_{4},0)$ via the S-duality, so they should also form the $4d$ adjoint massive vector multiplet. The whole KK tower of the $4d$ vector multiplet together may give the $6d$ $(2,0)$ tensor multiplet in adjoint representation of $U(N)$.

\section{Selfdual strings with the orientation covering a plane and the $M5$-$D3$ duality}

In this section, we will directly discuss the $M5$ branes and show that selfdual strings parallel to a given plane could offer the complete degrees of freedom on $M5$. The proposal is also supported by the $M5$-$D3$ duality, in which, the KK mode of $M5$ on $x_{4} \times x_{5}$ is dual to the $(p,q)$ open strings ending on $D3$ winding the transverse $x'^{45}$. When $M5$ branes are separated in $5d$ transverse space, on $D3$, 3-string junctions may form, which, in $M5$ picture, is the bound state of the unparallel selfdual strings.

\subsection{Selfdual strings parallel to the $45$ plane and the momentum mode of $M5$ on $x_{4} \times x_{5}$}

In previous discussion, we have seen that selfdual strings extending along all possible directions in $45$ plane may give the complete spectrum for a single $6d$ particle. The common eigenstates of $\hat{X}_{1}$ $\hat{X}_{2}$ $\hat{X}_{3}$ $\hat{P}_{4}$ $\hat{P}_{5}$,
\begin{equation}\label{MY}
	\Lambda = \left\{\delta (x_{1}-X_{1})\delta (x_{2}-X_{2})\delta (x_{3}-X_{3}) e^{iP_{4}x_{4}}e^{iP_{5}x_{5}}| \forall \: X_{1},X_{2},X_{3},P_{4},P_{5}  \right\}
\end{equation}
may be the suitable bases, the superposition of which can give a $6d$ particle localized in $(X_{1},X_{2},X_{3},X_{4}, X_{5})$. Although the position eigenstates can also be obtained, the basic excitations are $[i,j]$ selfdual strings other than the $[i,j]$ particles. Since it is (\ref{MY}) other than the position eigenstates that is naturally realized, the KK modes in this theory may tell us more than the KK modes in theories with point-like excitations.

Until now, our discussion is only restricted to coincident $M5$ branes. When $\vec{v}_{ij}\neq 0$, the selfdual string carrying $(P_{4},P_{5})$ momentum are massive. Consider the $[i,j]$ selfdual strings with the length and the orientation characterized by the vector $(qR_{4},pR_{5})$ in $45$ plane. $p$ and $q$ are co-prime, so the selfdual string only winds $x_{4} \times x_{5}$ once. In $123$ space, the string is localized at a point. The Wilson surface in $x_{4} \times x_{5}$ is trivial. Nevertheless, each $[i,j]$ string can still effectively pick up the background 2-form field $B_{45} = k/( 2 \pi R_{4}R_{5})$, $\forall \: k\in \textbf{N}$. $\forall \: m,n \in \textbf{Z}$, $\exists \: k,p,q$, $m= kq$, $n=kp$. In $45$ plane, the $[i,j]$ string will get the definite transverse momentum $(n/R_{4},-m/R_{5})$, thus could be taken as the plane wave $e^{i(\frac{nx_{4}}{R_{4}}-\frac{mx_{5}}{R_{5}})}$. If the momentum in $123$ space is $(P_{1},P_{2},P_{3})$, the energy will be
\begin{equation}\label{YMM}
	E = \sqrt{P^{2}_{1}+ P^{2}_{2} + P^{2}_{3}+ \frac{n^{2}}{R^{2}_{4}}+ \frac{m^{2}}{R^{2}_{5}} + 4 \pi^{2} |\vec{v}_{ij}|^{2}(q^{2}R^{2}_{4}+ p^{2}R^{2}_{5})}, 
\end{equation} 
since the $[i,j]$ selfdual string has the rest mass $2 \pi |\vec{v}_{ij}|\sqrt{q^{2}R^{2}_{4}+ p^{2}R^{2}_{5}}$.

Notice that there is an ambiguity for the mass of the zero mode in $4d$. With $k=0$, any $(qR_{4},pR_{5})$ string can be the $4d$ zero mode with mass $2 \pi |\vec{v}_{ij}|\sqrt{q^{2}R^{2}_{4}+ p^{2}R^{2}_{5}}$. However, the zero mode is unique. In the dual $D3$ picture, there are unwrapped $[i,j]$ $(p,q)$ strings with the length $|\vec{v}_{ij}|$. One may choose one possible $(qR_{4},pR_{5})$/$(p,q)$ as the zero mode. A particular S-frame is selected in this way, while in other S-frames, all $(qR_{4},pR_{5})$ can get the chance to act as the zero mode. For the point-like excitations, the $5d$ momentum can uniquely fix the state; however, for the line-like excitations, with the given $5d$ momentum, the selfdual string orientations can still vary in a $4d$ space orthogonal to the momentum. If the selfdual string orientations are restricted to a plane, a one-to-one correspondence between the momentum and state may be realized except for the momentums orthogonal to that plane. So, the fixing of the S-frame is necessary. As is shown in later discussions, the $M5$-$D3$ duality also intrinsically involves the selection of the S-frame.

For $M5$ compactified on $x_{5}$, the $P_{5}$ zero mode should be the state with the zero momentum along $x_{5}$, localized in $1234$ space. Selfdual string extending along $x_{5}$ is the only one meeting the requirement, and so, in Coulomb branch, the mass of the W-bosons is $2 \pi |\vec{v}_{ij}| R_{5}$ without the ambiguity. Notice that there is a distinction between the little string theory and the theory on $M5$ branes. For little string theory with $x_{5}$ compactified, there are momentum mode and the winding mode. The momentum mode is carried by the closed string. Although the string has the finite tension, the mass of the momentum mode is still $m/R_{5}$, because the string without winding $x_{5}$ can shrink to point thus has the zero mass and has no contribution to the energy. On the other hand, for $6d$ $(2,0)$ theory with $x_{5}$ compactified, the zero mode is the $5d$ SYM field. In Coulomb branch, the mass of the $[i,j]$ zero mode is $2 \pi |\vec{v}_{ij}|R_{5}$ other than $0$, since the zero mode is actually the selfdual string winding $x_{5}$ once. There is no way to get rid of the lowest winding mode, because we do not have the closed charged selfdual string, while the straight selfdual string localized in $1234$ space must extend along $x_{5}$.

The direct study of the $6d$ $(2,0)$ theory is difficult. The $x_{5}$ compactification will give the $5d$ massive tensor multiplet, which is also not quite accessible. The $4d$ KK modes upon the compactification on $x_{4} \times x_{5}$ are relatively easy to study. Moreover, the previous discussion indicates that (\ref{MY}) might be the suitable bases to consider the $6d$ theory, so in the following, we will focus on the $4d$ KK modes arising from the $6d$ theory.

\subsection{The $M5$-$D3$ duality}

Actually, $M5$ with $x_{4}$ and $x_{5}$ compactified in $11d$ spacetime is dual to $D3$ with one transverse dimension $x'^{45}$ compactified in $10d$ spacetime. Just as $M5$ with $x_{5}$ compactified in M theory is dual to $D4$ in type IIA string theory with $x_{5}$ being the M theory circle, $M5$ with $x_{4}$ and $x_{5}$ compactified in M theory is dual to $D3$ in type IIB string theory with $x'^{45}$ compactified. If the radii of $x_{4}$ and $x_{5}$ are $R_{4}$ and $R_{5}$, the radius of $x'^{45}$ will be $R'_{45} =1/(2 \pi R_{4}T_{F1}) =1/(2 \pi R_{5}T_{D1}) $. The five transverse dimensions of $M5$, $x^{I}$, are dual to the rest five transverse dimensions of $D3$, $x'^{I}$. $I=6 \cdots  10$. If the scalar fields on $M5$ branes get the vacuum expectation value $\Phi^{I}_{i}$ with $i=1\cdots N$, $D3$ branes will be separated along $x'^{I}$ with the transverse positions 
\begin{equation}
	X'^{I}_{i} = 2\pi R_{5}\Phi^{I}_{i}/T_{F1} = 2\pi R_{4}\Phi^{I}_{i}/T_{D1}. 
\end{equation}
$x_{4}\times x_{5}$ is dual to $x'^{45}$. If the $B_{45}$ on $M5$ branes gets the vacuum expectation value $B_{45i}$, $D3$ branes will be separated along $x'^{45}$ with the transverse positions 
\begin{equation}
	X'^{45}_{i} = 2\pi R_{5}B_{45i}/T_{F1} = 2\pi R_{4}B_{45i}/T_{D1}.
\end{equation} 
If the $B_{\mu 5}$ for $\mu=1,2,3$ on $M5$ branes gets the vacuum expectation value $B_{\mu 5 i}$, the gauge field $A_{\mu i}$ on $D3$ branes will get the vacuum expectation value 
\begin{equation}\label{S}
	A_{\mu i} = 2\pi R_{5}B_{\mu 5 i}.
\end{equation}
(\ref{S}) indicates that the $(0,0)$ mode of $M5$ is dual to the $(1,0)$ string on $D3$ with the winding number $0$. A particular S-frame is selected.

For T-duality, $Dp$ with the longitudinal $x_{p}$ compactified is dual to $D_{p-1}$ with the transverse $x'^{p}$ compactified, with $A_{p}$ converted to $X^{p}$, the momentum mode along $x_{p}$ transformed to the winding mode along $x'^{p}$. For $M5$, we have $B_{\mu\nu}$ instead of $A_{\mu}$, so two longitudinal dimensions $x_{4}$ $x_{5}$ are transformed to one transverse dimension $x'^{45}$, while the $(P_{4},P_{5})$ momentum modes become the winding modes of the $(p,q)$ strings along $x'^{45}$ for all co-prime $p$ $q$.

The $M5$-$D3$ duality requires that the both sides have the same degrees of freedom. Especially, the three vector fields and one scalar field on $D3$ are dual to the four 2-form fields $B_{\mu 5}$ on $M5$. The rest 2-form fields on $M5$ have no counterpart thus could be neglected. This is consistent with the self-duality condition on $M5$. Especially, for $M5$ compactified on $x_{4} \times x_{5}$,
\begin{equation}
	B_{\mu \nu } (x_{4},x_{5},\vec{x}) = \frac{1}{2\pi \sqrt{R_{4}R_{5}}}\sum_{n,m}e^{i(nx_{4}/R_{4}+mx_{5}/R_{5})}B_{\mu \nu }^{(n,m)} (\vec{x}). 
\end{equation}
The zero mode has no winding number around $x'^{45}$. $B_{4 5 }^{(0,0)} (\vec{x}) \rightarrow X^{4 5 (0,0)} (\vec{x})$, $B_{i 5 }^{(0,0)} (\vec{x}) \rightarrow A_{i}^{(0,0)} (\vec{x})$, where $i = 1,2,3$. The rest $B_{\mu \nu }^{(0,0)}$ could be neglected. The bosonic degrees of freedom are $6+2=8$. The higher mode has the nonzero $x'^{45}$ winding number, so there is no $X^{4 5 (n,m)}$ for $m,n \neq 0$. $B_{i 5 }^{(n,m)} (\vec{x}) \rightarrow A_{i}^{(n,m)} (\vec{x})$, where $i = 1,2,3$. $B_{4 5 }^{(n,m)} (\vec{x})$ together with $A_{i}^{(n,m)} (\vec{x})$ gives $4-1=3$ gauge degrees of freedom, so the total bosonic degrees of freedom are still $5+3=8$.

The $[i,j]$ $(qR_{4},pR_{5})$ selfdual string on $M5$ is dual to the $[i,j]$ $(p,q)$ string on $D3$. If $x_{4}$ and $x_{5}$ are compact, $x'_{45}$ will also be compact, so even if $B_{45} =0$, the covering space of $x'_{45}$ will still have $N$ coincident $D3$ branes distributed with the period $2 \pi R'_{45}$. $\forall \:i,j$, we have the $[i,j]$ $(p,q)$ string connecting the $i_{th}$ and the $j_{th}$ $D3$ branes with the length $2 \pi k R'_{45}$, corresponding to the $[i,j]$ $(qR_{4},pR_{5})$ selfdual string coupling with the 2-form field $B_{45} = k/(2 \pi R_{4}R_{5})$, getting the momentum $(kp/R_{4}, -kq/R_{5})$. If the other transverse fields on $D3$ also get the vacuum expectation value, the mass of the $[i,j]$ $(p,q)$ string will be 
\begin{equation}
	M= 2 \pi \sqrt{q^{2}R^{2}_{4}+p^{2}R^{2}_{5}}\sqrt{(\frac{k}{2 \pi R_{4}R_{5}})^{2}+|\vec{\Phi}_{ij}|^{2}},
\end{equation}
which is the same as the energy of the $[i,j]$ $(qR_{4},pR_{5})$ selfdual string.

The $(kp/R_{4}, -kq/R_{5})$ momentum mode is dual to the $[i,j]$ $(p,q)$ string winding $x'^{45}$ $k$ times. The $(p,q)$ string with all possible winding numbers gives a $5d$ SYM theory whose basic excitations are $(p,q)$ strings. In this way, the $4d$ KK modes are equivalent to a series of $5d$ SYM fields labeled by $(p,q)$ with $p$ and $q$ co-prime. The $(p,q)$ $5d$ SYM fields, when lifted to $6d$, are translation invariant along the $(qR_{4},pR_{5})$ direction. They are the fields related with the $(qR_{4},pR_{5})$ selfdual strings. With all $(p,q)$ included, the selfdual string orientation then covers the whole $45$ space.

\subsection{The 3-string junctions on $D3$ and $M5$}

When $\vec{v}_{i}=0$, all $D3$ branes are separated along a straight line, so the possible BPS states are still the original $1/2$ BPS states. To get the new states, the vacuum expectation values of the five scalar fields on $M5$ branes must be turned on. In $D3$ brane picture, $D3$'s then appear as $N$ arbitrary points in the $5d$ transverse space orthogonal to $x'^{45}$. The only possible new BPS states are $1/4$ BPS 3-string junctions, which are also the bound states of the $[i,j]$ $(p,q)$ string and the $[j,k]$ $(r,s)$ string. On $M5$ side, the 3-string junction is the bound state of the $[i,j]$ $[j,k]$ selfdual strings each carrying the transverse momentum $(kp/R_{4},-kq/R_{5})$ and $(hr/R_{4},-hs/R_{5})$\footnote{See \cite{113} for another discussion of the string junctions on $M5$ and $D4$.}.

We are interested with the $N$ coincident $M5$ branes, since in that case, the states can be massless in six dimensional sense thus will contribute to the entropy. We have seen that the bound states of the KK mode cannot give the new degrees of freedom, but we haven't considered the bound states of the KK mode and the zero mode, which, for example, can be taken as the tensionless $[i,j]$ $(0,R_{5})$ string. In $D3$ brane picture, that is the massless $[i,j]$ $(1,0)$ string, which may form the $1/4$ BPS threshold bound state with any $[j,k]$ $(p,q)$ strings with the length $2 \pi k R'_{45}$. In $6d$, they are the $[i,j]$ $(0,R_{5})$ and the $[j,k]$ $(qR_{4},pR_{5})$ tensionless selfdual strings located at the same point in $123$ space. The former has the zero momentum in $45$ plane, while the later carries the transverse momentum $(kp/R_{4},-kq/R_{5})$. A potential problem is that the threshold bound state may decay. If they do decay, then there will be no three indexed BPS states on $M5$.

\section{The interaction of the $6d$ $(2,0)$ theory seen from its KK modes on $x_{4} \times x_{5}$}

We now turn to the $6d$ $(2,0)$ theory on $M5$ branes. The 3-algebra valued $(2,0)$ tensor multiplet with the constant vector $C_{\mu}$ proposed in \cite{3} is the natural framework to describe the selfdual strings. Selfdual strings parallel to the $45$ plane give the fields $f_{(\theta)}(x_{\mu})$ with $C_{\mu}(\theta)\partial^{\mu}f_{(\theta)}(x_{\mu}) = 0$ for $C_{\mu}(\theta) = \cos \theta \delta^{4}_{\mu}+ \sin \theta \delta^{5}_{\mu}$, which altogether are equivalent to the $6d$ field. $f_{(\theta)}(x_{\mu})$ is in the adjoint representation of $U(N)$. Couplings like $Tr[f_{(\theta_{1})}\cdots f_{(\theta_{n})}]$ do not exist unless $\theta_{1} = \cdots =\theta_{n}$, because the bound state of two unparallel selfdual strings is the string junction other than another selfdual string. The quantization of the 3-string junction gives the $(2,1)$ multiplet $g_{(\theta_{1},\theta_{2},\theta_{3})}(x_{\mu})$ in $N \times N \times N$ and $\bar{N} \times \bar{N} \times \bar{N}$ representation of $U(N)$, which may couple with the vector multiplet $f_{(\theta)}(x_{\mu})$ as long as $\theta = \theta_{1},\theta_{2},\theta_{3}$. On coincident $M5$ branes, $g_{(\theta_{1},\theta_{2},\theta_{3})}(x_{\mu})$ reduces to $g_{(\theta)}(x_{\mu})$ subject to the constraint $C_{\mu}(\theta)\partial^{\mu}g_{(\theta)}(x_{\mu}) = 0$, giving a $6d$ field.

\subsection{$6d$ field decomposed into the $\theta$-parameterized $5d$ fields}

Recall that in \cite{3}, the equations of motion for the 3-algebra valued $(2,0)$ tensor multiplet involve a constant vector field $C_{\mu}$, where $\mu = 0 \cdots 5$, giving a direction along which all of the fields are required to be translation invariant. The theory with the fixed $C_{\mu}$ describes the selfdual string extending along it. The selfdual string has the zero momentum along $C_{\mu}$ but may get the arbitrary momentum along the four transverse dimensions, so the theory describing it is just the $5d$ $U(N)$ SYM theory, which is the reduction of the $6d$ $(2,0)$ theory along $C_{\mu}$. Moreover, as the zero mode along $C_{\mu}$, the field configurations of the $5d$ SYM theory on $R^{4}$ should carry the zero instanton number (Pontryagin number) thus are topologically trivial on the equivalent $S^{4}$. To recover the full $6d$ theory, we need the selfdual strings with the orientations covering all directions in a plane, which, for definiteness, is taken as the $45$ plane. Correspondingly, $C_{\mu}$ is replaced by $C_{\mu}(\theta) = \cos \theta \delta^{4}_{\mu}+ \sin \theta \delta^{5}_{\mu}$, while the original fields $f(x_{\mu})$ now become $f(\theta, x_{\mu})$ still with the constraint $C_{\mu}(\theta)\partial^{\mu}f(\theta, x_{\mu}) = 0$, giving rise to a $6d$ field.

Suppose the $U(N)$ $6d$ $(2,0)$ tensor multiplet field configuration is given. For simplicity, consider the scalar fields $X^{I}(x_{m}, x_{4}, x_{5} )$, where $m=0,1,2,3$, $I = 6,7,8,9,10$. 
\begin{equation}\label{9870}
	\int dx_{\theta} \: X^{I}(x_{m}, \cos \theta x_{\theta} + \sin \theta y_{\theta}, - \sin \theta x_{\theta} + \cos  \theta y_{\theta} ) = \Phi^{I} (x_{m}, \theta, y_{\theta})
\end{equation}
is the scalar field in the $5d$ SYM theory related with $\theta$. $X^{I}$ and $\Phi^{I}$ have the scaling dimensions $2$ and $1$ respectively. $\Phi^{I}$ is the zero mode of $X^{I}$ along $C_{\mu}(\theta)$. Also, notice that
\begin{equation}
	\int dx_{\theta}dy_{\theta} \: X^{I}(x_{m}, \cos \theta x_{\theta} + \sin \theta y_{\theta}, - \sin \theta x_{\theta} + \cos  \theta y_{\theta} ) = \phi^{I} (x_{m})  
\end{equation}
is independent of $\theta$. $\phi^{I}$ is the zero mode of $\Phi^{I} (x_{m}, \theta, y_{\theta})$ in the $4d$ spacetime. All of the $5d$ SYM theories share the same zero mode in $4d$, because the $6d$ theory has the unique zero mode in $4d$. The vector field $A$ and the spinor field $\eta$ with the scaling dimensions $1$ and $3/2$ in $5d$ SYM theory could be constructed in the similar way from the $6d$ 2-form field $B$ and the spinor field $\Psi$ with the scaling dimensions $2$ and $5/2$. Since the integration is carried out along a particular direction, more precisely, the original scalar fields, 2-form field, and the spinor field are converted into the vector fields, vector field, and the spinor-vector field respectively. 

One may also want to reconstruct $X^{I}$ from $\Phi^{I}$.
\begin{equation}\label{qwe}
	X^{I}(x_{m}, x_{4}, x_{5} ) = \int dp_{4} dp_{5} \: e^{i(p_{4}x_{4}+p_{5}x_{5})}\phi^{I}_{(p_{4},p_{5})}(x_{m}).
\end{equation}
If $x_{4}$ and $x_{5}$ are compact, $p_{4}=k\bar{p}_{4}/R_{4}$, $p_{5}=-k\bar{p}_{5}/R_{5}$, $(\bar{p}_{4},\bar{p}_{5})$ is the co-prime pair,
\begin{equation}
	X^{I}(x_{m}, x_{4}, x_{5} ) = \sum_{(\bar{p}_{4},\bar{p}_{5})}\sum_{k} \: e^{ik(\frac{\bar{p}_{4}x_{4}}{R_{4}}-\frac{\bar{p}_{5}x_{5}}{R_{5}})}\phi^{I}_{(\bar{p}_{4},\bar{p}_{5};k)}(x_{m})=\sum_{(\bar{p}_{4},\bar{p}_{5})} \Phi^{I}(x_{m},\bar{p}_{4}R_{5}x_{4}-\bar{p}_{5}R_{4}x_{5} ).
\end{equation}
$\Phi^{I}(x_{m},\bar{p}_{4}R_{5}x_{4}-\bar{p}_{5}R_{4}x_{5} )$ is the discrete version of $\Phi^{I}$. In continuous limit,
\begin{equation}
	X^{I}(x_{m}, x_{4}, x_{5} ) = \int d\theta dp_{\theta}p_{\theta} \;  e^{ip_{\theta}(-\sin \theta x_{4}+\cos \theta x_{5})}\phi^{I}_{(\theta;p_{\theta})}(x_{m})=\int d\theta \; \tilde{\Phi}^{I}(x_{m},-\sin \theta x_{4}+\cos \theta x_{5}).
\end{equation}
However, $\tilde{\Phi}^{I}$ is not the $\Phi^{I}$ in (\ref{9870}). The latter is 
\begin{equation}
	\Phi^{I} (x_{m},-\sin \theta x_{4}+\cos \theta x_{5}) =   \int  dp_{\theta} \;  e^{ip_{\theta}(-\sin \theta x_{4}+\cos \theta x_{5})}\phi^{I}_{(\theta;p_{\theta})}(x_{m})
\end{equation}
with $p_{\theta}$ left out in the integral. $X^{I}$ is only the direct superposition of $\tilde{\Phi}^{I}(\theta)$, which is not the zero mode in $C_{\mu}(\theta)$ direction. Nevertheless, $\left\{ \tilde{\Phi}^{I}(\theta)\: |\: \forall \theta \in [0,\pi) \right\}$ and $\left\{\Phi^{I}(\theta)\:|\: \forall \theta \in [0,\pi) \right\}$ are equivalent bases, so it is indeed possible to reconstruct $X^{I}$ from $\Phi^{I}$.

\subsection{The coupling of the selfdual strings and the 3-string junctions}

We now have a series of $\theta$-parameterized $5d$ $U(N)$ SYM theories, which is effectively a $6d$ theory with $5$ scalars, $3$ gauge degrees of freedom and $8$ fermionic degrees of freedom. It may at least exhaust the $1/2$ BPS field content of the $6d$ $(2,0)$ theory. The next problem is the interaction. Fields belong to the same $5d$ SYM theory have the standard SYM coupling among themselves. It is also necessary to consider the couplings involving fields in different $5d$ SYM theories. Actually, the $6d$ SYM theory could also be decomposed in this way, while the local interactions in the original $6d$ theory induce the couplings among the $5d$ theories labeled by different $\theta$.

To see this coupling more explicitly, we'd better decompose the $6d$ fields into the $4d$ KK modes. For scalars, the decomposition is as that in (\ref{qwe}). Similarly, for the $6d$ SYM fields such as the scalars $Y^{L}(x_{m}, x_{4}, x_{5} )$, $L = 6,7,8,9$, there is also 
\begin{equation}
	Y^{L}(x_{m}, x_{4}, x_{5} ) = \int dp_{4} dp_{5} \: e^{i(p_{4}x_{4}+p_{5}x_{5})}\varphi^{L}_{(p_{4},p_{5})}(x_{m}). 
\end{equation}
The two-field coupling $Y^{L}Y^{L'}$ gives
\begin{equation}
	\int dx_{4} dx_{5} \: Y_{ij}^{L}Y_{ji}^{L'} = \int dp_{4} dp_{5} \: \varphi^{L}_{(p_{4},p_{5})ij}(x_{m})\varphi^{L'}_{(-p_{4},-p_{5})ji}(x_{m}), 
\end{equation}
the three-field coupling $Y^{L}Y^{L'}Y^{L''}$ gives 
\begin{equation}\label{label}
	\int dx_{4} dx_{5} \: Y_{ij}^{L}Y_{jk}^{L'}Y_{ki}^{L''} = \int dp_{4} dp_{5} dq_{4} dq_{5} \: \varphi^{L}_{(p_{4},p_{5})ij}(x_{m})\varphi^{L'}_{(q_{4},q_{5})jk}(x_{m})\varphi^{L''}_{(-p_{4}-q_{4},-p_{5}-q_{5})ki}(x_{m}), 
\end{equation}
and similarly for the $n$-field coupling. In the dual $D3$ brane picture, $\varphi^{L}_{(p_{4},p_{5})ij}(x_{m})$ corresponds to the F-string connecting the $i_{th}$ and the $j_{th}$ $D3$ branes represented by the vector $(p_{4},p_{5})$ in transverse space. The above coupling is possible because the bound state of the $[i,j]$ $(p_{4},p_{5})$ F-string and the $[j,k]$ $(q_{4},q_{5})$ F-string is the $[i,k]$ $(p_{4}+q_{4},p_{5}+q_{5})$ F-string. The conclusion also holds in Coulomb branch. $\varphi^{L}_{(p_{4},p_{5})ij}(x_{m})$ then corresponds to the F-string represented by the vector $(p_{4},p_{5}, \vec{v}_{ij})$ in transverse space. 
\begin{equation}
	(p_{4},p_{5}, \vec{v}_{ij})+(q_{4},q_{5}, \vec{v}_{jk}) = (p_{4}+q_{4},p_{5}+q_{5}, \vec{v}_{ik}). 
\end{equation}
On the other hand, for fields in tensor multiplet, such as $X^{I}$, the two-field coupling $X^{I}X^{I'}$ is indeed
\begin{equation}\label{sti}
		\int dx_{4} dx_{5} \: X_{ij}^{I}X_{ji}^{I'} = \int dp_{4} dp_{5} \: \phi^{I}_{(p_{4},p_{5})ij}(x_{m})\phi^{I'}_{(-p_{4},-p_{5})ji}(x_{m}), 
\end{equation}
but the three-field coupling and the $n$-field coupling cannot take the similar form as (\ref{label}). On $D3$ branes, $\phi^{I}_{(p_{4},p_{5})ij}(x_{m})$ corresponds to the $[i,j]$ $(p_{4},p_{5})$ string\footnote{More accurately, it is the $[i,j]$ $(\bar{p}_{4},\bar{p}_{5})$ string winding $x'^{45}$ $k$ times. $p_{4} = k \bar{p}_{4}/R_{4} $, $p_{5}=- k \bar{p}_{5}/R_{5} $, $\bar{p}_{4}$ and $\bar{p}_{5}$ are co-prime. For simplicity, we just denote it by $(p_{4},p_{5})$.}. When $(p_{4},p_{5}) \propto (q_{4},q_{5})$, the bound state of the $[i,j]$ $(p_{4},p_{5})$ string and the $[j,k]$ $(q_{4},q_{5})$ string is still the $[i,k]$ $(p_{4}+q_{4},p_{5}+q_{5})$ string, so, the coupling like (\ref{label}) is possible. $\phi^{I}_{(p_{4},p_{5})ij}(x_{m})$ and $\phi^{I'}_{(q_{4},q_{5})jk}(x_{m})$ belong to the same $5d$ SYM theory with $\theta = -\arctan (p_{4}/p_{5})$. However, for unparallel $(p_{4},p_{5})$ and $(q_{4},q_{5})$, the bound state will be the 3-string junction other than the single string\footnote{The bound state exists only when the $[i,j]$ $[j,k]$ strings have the suitable mass. Here, we just assume so.}. The similar problem also exists for the $5d$ massive tensor multiplet. The KK modes in $4d$ could be represented by the $[i,j]$ $(p_{4},p_{5})$ strings with the fixed $p_{5}$ but all possible $p_{4}$. $(p_{4},p_{5})$ and $(p'_{4},p_{5})$ are not parallel unless $p_{4} = p'_{4}$. So, if we concentrate on a single kind of the selfdual strings, the theory will be the $5d$ SYM theory; if we consider the selfdual strings with the different orientations, the theory will involve the tensor multiplet, for which, the interaction is not the standard SYM type.

Then the problem reduces to the coupling between $\phi_{(p_{4},p_{5})ij}(x_{m})$ and $\phi'_{(q_{4},q_{5})jk}(x_{m})$ for the unparallel $(p_{4},p_{5})$ and $(q_{4},q_{5})$. The bound state of the $[i,j]$ $(p_{4},p_{5})$ string and the $[j,k]$ $(q_{4},q_{5})$ string is the 3-string junction other than the traditional $[i,k]$ $(P_{4},P_{5})$ string. Unlike the $6d$ SYM theory, we now get more states and should also quantize them. A given 3-string junction is characterized by the charge vector $\textbf{v}_{e} = (r_{4},s_{4}, t_{4})$ and $\textbf{v}_{m} = (r_{5},s_{5}, t_{5})$, for which, no common divisor exists. $r$ $s$ $t$ are related with the $i$ $j$ $k$ branes, while the rest $N-3$ branes are neglected.
\begin{equation}
	r_{4}+s_{4}+ t_{4} = r_{5}+s_{5}+ t_{5}=0. 
\end{equation}
In $x^{45}$, $v^{45}_{ij}=2 \pi k R'_{45}$, $v^{45}_{jk}=2 \pi h R'_{45}$. In transverse space, we may also have $\vec{v}_{ij}$ and $\vec{v}_{jk}$, which will make the string junction massive. The total momentum of the 3-string junction is
\begin{equation}
	(P_{4},P_{5}) =(\frac{kr_{4}-h t_{4}}{R_{4}},\frac{-kr_{5}+h t_{5}}{R_{5}}) = (p_{4}+q_{4},p_{5}+q_{5}).  
\end{equation}
where $(p_{4},p_{5})=(kr_{4}/R_{4}, -kr_{5}/R_{5})=(k\bar{p}_{4}/R_{4},-k\bar{p}_{5}/R_{5})$, $(q_{4},q_{5})=(-h t_{4}/R_{4},h t_{5}/R_{5})=(h\bar{q}_{4}/R_{4},- h\bar{q}_{5}/R_{5})$. $\bar{p}_{4}$ and $\bar{p}_{5}$, $\bar{q}_{4}$ and $\bar{q}_{5}$ are not necessarily co-prime now. For $(\bar{p}_{4},\bar{p}_{5})\propto (\bar{q}_{4},\bar{q}_{5})$, $(P_{4},P_{5})\propto (\bar{q}_{4},\bar{q}_{5})\propto (\bar{p}_{4},\bar{p}_{5})$, while for the unparallel $(\bar{p}_{4},\bar{p}_{5})$ and $(\bar{q}_{4},\bar{q}_{5})$, $k$ and $h$ may generate two dimensional momentum. Especially, if the $SL(2,\textbf{Z})$ invariant intersection number \cite{7} $I = t_{5}r_{4}-t_{4}r_{5}=\pm1$, $(P_{4},P_{5})$ can cover all of $(n/R_{4},m/R_{5})$; otherwise, it can only cover $(nI/R_{4},mI/R_{5})$. We will use $(\bar{p}_{4}, \bar{p}_{5})$, $(\bar{q}_{4}, \bar{q}_{5})$ and $(P_{4},P_{5})$ to denote the 3-string junction. When $R_{4},R_{5}\rightarrow \infty$, $R_{4}/ R_{5}$ is indefinite, so the 3-string junction is denoted by $(\theta_{1}, \theta_{2}, \theta_{3})$. For the given $\vec{v}_{ij}$ and $\vec{v}_{jk}$, the 3-string junction exists only when $(P_{4},P_{5})$ satisfies some particular condition 
\begin{equation}\label{wi}
	h_{(\bar{p}_{4}, \bar{p}_{5}, \bar{q}_{4}, \bar{q}_{5}, \vec{v}_{ij}, \vec{v}_{jk})}(P_{4},P_{5})>0,
\end{equation}
so the 3-string junction cannot have the arbitrary momentum in $45$ plane. The selfdual string in $45$ plane can only carry the $1d$ transverse momentum. Here, the bound state of two unparallel selfdual strings can carry the $2d$ momentum, but this momentum cannot cover the $2d$ space.

Fields arising from the quantization of the 3-string junctions can then be denoted by $\phi_{(\vec{r}, \vec{s}, \vec{t}; P_{4},P_{5})ijk}(x_{m})$ or $\phi_{(\theta_{1},\theta_{2},\theta_{3}; P_{4},P_{5})ijk}(x_{m})$ in decompactification limit. $\vec{r}+ \vec{s}+ \vec{t}=0$. The corresponding $6d$ field is
\begin{equation}\label{witt}
		X_{(\vec{r}, \vec{s}, \vec{t})ijk}(x_{m}, x_{4}, x_{5}) =
\sum_{P_{4},P_{5}} \;
e^{i(P_{4}x_{4}+P_{5}x_{5})}\phi_{(\vec{r}, \vec{s}, \vec{t}; P_{4},P_{5})ijk}(x_{m}),  
\end{equation}
or  
\begin{equation}\label{wit}
		X_{(\theta_{1},\theta_{2},\theta_{3})ijk}(x_{m}, x_{4}, x_{5}) =
\int dP_{4} dP_{5} \;
e^{i(P_{4}x_{4}+P_{5}x_{5})}\phi_{(\theta_{1},\theta_{2},\theta_{3}; P_{4},P_{5})ijk}(x_{m}).   
\end{equation}
In polar coordinate, (\ref{wit}) could also be written as
\begin{eqnarray}\label{wittt}
\nonumber 	&& 	X_{(\theta_{1},\theta_{2},\theta_{3})ijk}(x_{m}, \theta, \rho)   \\  &= & \sin (\theta_{2}-\theta_{1})\int dp_{\theta_{1}} dp_{\theta_{2}} \;
e^{i\rho[\sin (\theta-\theta_{1})p_{\theta_{1}} - \sin (\theta-\theta_{2}) p_{\theta_{2}}  ]}\phi_{(\theta_{1},\theta_{2},\theta_{3}; p_{\theta_{1}}, p_{\theta_{2}})ijk}(x_{m}).    \  
\end{eqnarray}
In terms of $k$ and $h$, (\ref{witt}) becomes 
\begin{equation}\label{wii}
	X_{(\vec{r}, \vec{s}, \vec{t})ijk}(x_{m}, x_{4}, x_{5})=\sum_{k,h} \;
e^{i[k(\frac{r_{4}x_{4}}{R_{4}}- \frac{r_{5}x_{5}}{R_{5}})-h(\frac{t_{4}x_{4}}{R_{4}}- \frac{t_{5}x_{5}}{R_{5}})]}\phi_{(\vec{r}, \vec{s}, \vec{t}; k,h)ijk}(x_{m}). 
\end{equation}
In (\ref{witt})-(\ref{wii}), $(P_{4},P_{5})$ is in the range specified by (\ref{wi}).

With the fields related with strings as well as the string junctions, we can consider the possible couplings among them. First, the bound state of the $[i,j]$ $(r_{4},r_{5})$ string and the $[j,k]$ $(-t_{4},-t_{5})$ string, or the $[j,k]$ $(s_{4},s_{5})$ string and the $[k,i]$ $(-r_{4},-r_{5})$ string, or the $[k,i]$ $(t_{4},t_{5})$ string and the $[i,j]$ $(-s_{4},-s_{5})$ string is the $[i,j,k]$ 3-string junction with $\textbf{v}_{e} = (r_{4},s_{4}, t_{4})$ and $\textbf{v}_{m} = (r_{5},s_{5}, t_{5})$. The momentum of the 3-string junction is the sum of the two individual strings. $2+2 \rightarrow 3$. 
\begin{equation}
	\phi_{(r_{4},r_{5}; k)ij}(x_{m}) \phi'_{(-t_{4},-t_{5}; h)jk}(x_{m}) \sim \phi''_{(\vec{r}, \vec{s}, \vec{t}; k,h)ijk}(x_{m}),   
\end{equation}
\begin{equation}
	\phi_{(s_{4},s_{5}; h)jk}(x_{m}) \phi'_{(-r_{4},-r_{5}; -k-h)ki}(x_{m}) \sim \phi''_{(\vec{r}, \vec{s}, \vec{t}; k,h)ijk}(x_{m}),   
\end{equation}
\begin{equation}
	\phi_{(t_{4},t_{5}; -k-h)ki}(x_{m}) \phi'_{(-s_{4},-s_{5}; k)ij}(x_{m}) \sim \phi''_{(\vec{r}, \vec{s}, \vec{t}; k,h)ijk}(x_{m}),   
\end{equation}
which could be derived from the coupling 
\begin{equation}
\int dx_{4}dx_{5} \; X_{(r_{4}, r_{5})ij}(x_{m}, x_{4}, x_{5})	X'_{( -t_{4}, -t_{5})jk}(x_{m}, x_{4}, x_{5})	X''_{(\vec{r}, \vec{s}, \vec{t})ijk}(x_{m}, x_{4}, x_{5}),
\end{equation}
with 
\begin{equation}\label{opop}
	X_{(r_{4}, r_{5})ij}(x_{m}, x_{4}, x_{5}) = \sum_{k} \; e^{ik(\frac{r_{4}x_{4}}{R_{4}}- \frac{r_{5}x_{5}}{R_{5}})}\phi_{(r_{4}, r_{5}; k)ij}(x_{m}), 
\end{equation}
\begin{equation}
	X'_{(-t_{4}, -t_{5})jk}(x_{m}, x_{4}, x_{5}) = \sum_{h} \; e^{-ih(\frac{t_{4}x_{4}}{R_{4}}- \frac{t_{5}x_{5}}{R_{5}})}\phi_{(-t_{4}, -t_{5}; h)jk}(x_{m}).  
\end{equation}
Next, consider the bound state of $\phi_{(u_{4}, u_{5}; g)li}(x_{m})$ or $\phi_{(u_{4}, u_{5}; -g)il}(x_{m})$ and $\phi'_{(\vec{r}, \vec{s}, \vec{t}; k,h)ijk}(x_{m})$. If $(u_{4}, u_{5})=(r_{4}, r_{5})$ or $(u_{4}, u_{5})=(-r_{4}, -r_{5})$, the bound state will still be the 3-string junction $\phi''_{(\vec{r}, \vec{s}, \vec{t}; k+g,h)ljk}(x_{m})$, $2+3 \rightarrow 3$; otherwise, it is a 4-string junction, $2+3 \rightarrow 4$. The situation is similar if $i$ is replaced by $j$ or $k$. The $2+3 \rightarrow 3$ type relation may give the couplings like
\begin{equation}
	X_{li}X'_{ijk}X''_{ljk}, \;\;\;\;X_{il}X'_{ijk}X''_{ljk}, \;\;\;\; X_{li}X'_{ijk}X''_{kn}X'''_{ljn}
\end{equation}
and so on.

The $2+2 \rightarrow 2$, $2+3 \rightarrow 3$ couplings could be realized as the matrix multiplication. Moreover, they can also be visualized as the junction of two 2-boundary-$M2$'s and the junction of one 2-boundary-$M2$ and one 3-boundary-$M2$ respectively. Therefore, they are more reasonable than the couplings like $2+2 \rightarrow 3$ and $2+3 \rightarrow 4$.

\subsection{The multiplet structure of the 3-string junctions}

We now have two sets of fields $f_{(r_{4}, r_{5})ij}(x_{m}, x_{4}, x_{5})$ and $	g_{(\vec{r}, \vec{s}, \vec{t})ijk}(x_{m}, x_{4}, x_{5})$, or alternatively, $f_{(\theta)ij}(x_{m}, \alpha, \rho)$ and $g_{(\theta_{1},\theta_{2},\theta_{3} )ijk}(x_{m},\alpha, \rho)$. $f_{(r_{4}, r_{5})ij}(x_{m}, x_{4}, x_{5})$ is translation invariant along the $(r_{5}R_{4}, r_{4}R_{5})$ direction. It is the previously discussed field satisfying the constraint $C_{\mu}(\theta)\partial^{\mu}f_{(\theta)ij}(x_{m}, \alpha, \rho)=0$. On the other hand, $g_{(\theta_{1},\theta_{2},\theta_{3})ijk}(x_{m},\alpha, \rho)$ is a $6d$ field without the constraint\footnote{(\ref{wi}) gives a restriction on the range of $k$ and $h$ in (\ref{wii}). Especially, if $k=0$ or $h=0$, $g_{(\theta_{1},\theta_{2},\theta_{3})ijk}$ is also translation invariant along one direction, as we will see later.}. $f_{(\theta_{1})}$, $f_{(\theta_{2})}$ and $f_{(\theta_{3})}$ may couple with each other through $g_{(\theta_{1},\theta_{2},\theta_{3})}$.

$f_{(\theta)ij}$ is a vector multiplet composed by the scalars $\Phi_{(\theta)}^{I}$, the vector $A_{(\theta)\mu}$ and the spinor $\eta_{(\theta)}$ with the scaling dimensions $1$, $1$ and $3/2$ respectively, coming from the $C_{\mu}(\theta)$ direction integration of the scalars $X^{I}$, the 2-form $B_{\mu\nu}$ and the spinor $\Psi$ with the scaling dimensions $2$, $2$ and $5/2$. As a $6d$ vector, $C^{\mu}(\theta)A_{(\theta)\mu} = 0$.

The field content of $g_{(\theta_{1},\theta_{2},\theta_{3})ijk}$ can be reconstructed from the $4d$ KK mode. The KK compactification of $g_{(\theta_{1},\theta_{2},\theta_{3})ijk}$ on $x_{4} \times x_{5}$ gives the $4d$ field $\phi_{(\vec{r}, \vec{s}, \vec{t};P_{4},P_{5})ijk}$, which, in $4d$ SYM theory, is related to the 3-string junction with the charge vector $\textbf{v}_{e} = (r_{4},s_{4}, t_{4})$ and $\textbf{v}_{m} = (r_{5},s_{5}, t_{5})$, having the total mass $P_{5}$ and the total electric charge $P_{4}$. The multiplet structure of $\phi_{(\vec{r}, \vec{s}, \vec{t};P_{4},P_{5})ijk}$ is $V_{4}\otimes V_{in}$, where $V_{4}$ is the vector supermultiplet coming from the free center-of-mass part, $V_{in}$ is the internal part determined by $\textbf{v}_{e}$ and $\textbf{v}_{m }$. For $\textbf{v}_{e} = (r_{4},s_{4},t_{4})$, $\textbf{v}_{m } = (1,0,-1)$, $ V_{in} = [|s_{4}|/2]\oplus[|s_{4}|/2-1/2]\oplus[|s_{4}|/2-1/2]\oplus[|s_{4}|/2-1]$, giving a total of $4|s_{4}|$ states \cite{6}. As the string web, it has $E_{ext}=3$ external points and $F_{int} = |s_{4}|$ internal points \cite{7}. If $\phi_{(\vec{r}, \vec{s}, \vec{t};P_{4},P_{5})ijk}$ is lifted into the $6d$ field $g_{(\vec{r}, \vec{s}, \vec{t})ijk}$, $V_{4}\otimes V_{in}$ will become $T_{6}\otimes V_{in}$, with $T_{6}$ the tensor supermultiplet from the center-of-mass part. So, $g_{(\vec{r}, \vec{s}, \vec{t})ijk}$ at least contains a tensor multiplet factor.

It is difficult to determine $V_{in}$ in decompactification limit. The simplest situation is $|s_{4}|=1$ with one internal point, and then $V_{in} = [1/2]\oplus [0]\oplus [0]$. Recall that for $1/2$ BPS states with the degeneracy of $2^{4}$, we have the $6d$ $(2,0)$ tensor multiplet $T_{6}$, whose KK modes along $x_{5}$ are the $5d$ massless vector multiplet $V_{5}$ and the $5d$ massive $(2,0)$ tensor multiplets $T_{5}$. The KK modes of $V_{5}$ and $T_{5}$ on $x_{4}$ are the $4d$ vector multiplets $V_{4}$. The massless limit of the $5d$ massive tensor multiplet is the $5d$ massless vector multiplet. For $1/4$ BPS states, in $4d$, we get $V_{4}\otimes ([1/2]\oplus [0]\oplus [0])$. The $V_{4} \otimes [1/2]$ part gives  
\vskip 5mm
\begin{center}\label{Mm}
\begin{tabular}{|c|c|c|c|c|c|c|c|}
\hline
$j$ & 3/2 & 1 & 1/2 & 0 & -1/2 & -1 & -3/2 \\
\hline
Degeneracy & 1 & 4 & 7 & 8 & 7 & 4 & 1\\
\hline
\end{tabular}
\end{center}
which, when combines with the rest two $V_{4}$, could be organized into $1$ spin-3/2 fermion, $6$ vectors, $14$ spin-1/2 fermions and $14$ scalars, forming the massive representation of the $4d$ ${\cal N}=4$ superalgebra with $2^{6}$ states. In massless limit, the bosonic part of $V_{4} \otimes ([1/2]\oplus [0]\oplus [0])$ is composed by $6$ vectors and $20$ scalars, with each vector containing two degrees of freedom. $V_{4}$, when lifted to $5d$ with $P_{5}=0$ or $P_{5}\neq 0$, becomes $V_{5}$ or $T_{5}$. The lifted $V_{4} \otimes ([1/2]\oplus [0]\oplus [0])$ could be naively denoted by $V_{5} \otimes ([1/2]\oplus [0]\oplus [0])$ and $T_{5} \otimes ([1/2]\oplus [0]\oplus [0])$, which are all complex now. Actually, one $V_{4} \otimes ([1/2]\oplus [0]\oplus [0])$ only gives the 3-string junction with one possible orientation; if the other orientation is taken into account, we will also get $2^{6} \times 2= 2^{7}$ states. The field content of $V_{5} \otimes ([1/2]\oplus [0]\oplus [0]) $ could be organized into $1$ spin-3/2 fermion, $6$ vectors, $13$ spin-1/2 fermions and $14$ scalars, while the field content of $T_{5} \otimes ([1/2]\oplus [0]\oplus [0]) $ could be organized into $2$ selfdual tensors, $1$ spin-3/2 fermion, $4$ vectors, $13$ spin-1/2 fermions and $10$ scalars. $T_{6} \otimes ([1/2]\oplus [0]\oplus [0])$ and $T_{5} \otimes ([1/2]\oplus [0]\oplus [0])$ have the same field content, forming the $6d$ massless $(2,1)$ multiplet and the $5d$ massive $(2,1)$ multiplet respectively. The $5d$ massive selfdual tensors and the $5d$ massive vectors, containing $3$ and $4$ degrees of freedom, become the $6d$ massless selfdual tensors and the $6d$ massless vectors, still with $3$ and $4$ degrees of freedom.

$T_{6} \otimes ([1/2]\oplus [0]\oplus [0])$ compactified on $x_{5}$ gives $V_{5} \otimes ([1/2]\oplus [0]\oplus [0])$ and $T_{5} \otimes ([1/2]\oplus [0]\oplus [0])$, which, when further compactified on $x_{4}$, becomes $V_{4} \otimes ([1/2]\oplus [0]\oplus [0])$. Just as $V_{5}$ is the massless limit of $T_{5}$, $V_{5} \otimes ([1/2]\oplus [0]\oplus [0])$ could also be taken as the massless limit of $T_{5} \otimes ([1/2]\oplus [0]\oplus [0])$. The $2$ massive $5d$ selfdual tensors become $2$ massless $5d$ vectors, while the $4$ massive $5d$ vectors become $4$ massless $5d$ vectors plus $4$ scalars. $T_{6} \otimes ([1/2]\oplus [0]\oplus [0])$, $T_{5} \otimes ([1/2]\oplus [0]\oplus [0]) $ and $V_{5} \otimes ([1/2]\oplus [0]\oplus [0]) $ are all complex, so the total states for each are $2^{7}$ other than $2^{6}$. Each multiplet will form the $N \times N \times N$ or $\bar{N}\times \bar{N}\times \bar{N}$ representation of $U(N)$, so they cannot be real, as the fields in adjoint representation do.

\subsection{The coupling among the vector multiplet and the $(2,1)$ multiplet}

$g_{(\theta_{1}, \theta_{2}, \theta_{3})ijk}$ is the $(2,1)$ multiplet composed by the scalars $X_{(\theta_{1}, \theta_{2}, \theta_{3})}$, the vectors $V_{(\theta_{1}, \theta_{2}, \theta_{3})}$, the 2-forms $B_{(\theta_{1}, \theta_{2}, \theta_{3})}$, the spin-1/2 fermions $\Psi_{(\theta_{1}, \theta_{2}, \theta_{3})}$ and the spin-3/2 fermions $\eta_{(\theta_{1}, \theta_{2}, \theta_{3})}$. In principle, the $6d$ $(2,0)$ theory can only contain the $(2,0)$ tensor multiplet, but now, the $(2,1)$ multiplet is also added. There will be the couplings between the $(2,1)$ multiplet and the vector multiplet arising from the reduction of the tensor multiplet along a particular direction. The incorporation of the $(2,1)$ multiplet into the scattering amplitude is also discussed in \cite{1111} for $M5$ compactified on $S^{1}$. It was shown that the $BB'A$ coupling is one of the possibilities. $B$ and $B'$ are the 2-forms in $5d$ massive $(2,1)$ multiplet, while $A$ is the zero mode vector in $5d$. In the following, we will only discuss $X$, $B$ and $\Psi$ with the scaling dimensions $2$, $2$, $5/2$ respectively, neglecting $V$ and $\eta$. The transverse indices of $X$, $B$ and $\Psi$ are dropped for simplicity, although $X$, $B$ and $\Psi$ are not the R-symmetry singlet.

Let us consider the possible dimension six couplings for these fields. For two-field couplings, there are 
\begin{equation}
	\partial X	\partial X, \;\;	\partial X_{(\theta_{1}, \theta_{2}, \theta_{3})}	\partial X^{*}_{(\theta_{1}, \theta_{2}, \theta_{3})},\;\; \bar{\Psi}\partial \Psi,\;\; \bar{\Psi}_{(\theta_{1}, \theta_{2}, \theta_{3})}\partial \Psi_{(\theta_{1}, \theta_{2}, \theta_{3})},\;\; \partial B	\partial B, \;\;	\partial B_{(\theta_{1}, \theta_{2}, \theta_{3})}	\partial B^{*}_{(\theta_{1}, \theta_{2}, \theta_{3})}. 
\end{equation}
$X$ $\Psi$ $B$ compose a $6d$ $(2,0)$ tensor multiplet in adjoint representation of $U(N)$, which is equivalent to $\Phi_{(\theta)}$ $A_{(\theta)}$ $\eta_{(\theta)}$ with all $\theta$ included. We do not have terms like $X_{ij}X'_{jk}X''_{ki}$, but the two-field couplings like $X_{ij}X'_{ji}$ are allowed. The tensor multiplet representation works well in free theory. The possible three-field couplings are 
\begin{equation}
A_{\theta_{a}}X_{(\theta_{1}, \theta_{2},\theta_{3})} \partial X^{*}_{(\theta_{1}, \theta_{2}, \theta_{3})}, \; A_{\theta_{a}} \bar{\Psi}_{(\theta_{1}, \theta_{2}, \theta_{3})} \Psi_{(\theta_{1}, \theta_{2}, \theta_{3})},\; \Phi_{\theta_{a}} \bar{\Psi}_{(\theta_{1}, \theta_{2}, \theta_{3})} \Psi_{(\theta_{1}, \theta_{2},\theta_{3} )},	\; A_{\theta_{a}}B_{(\theta_{1}, \theta_{2}, \theta_{3})}\partial B^{*}_{(\theta_{1}, \theta_{2}, \theta_{3})},  
\end{equation}
where $a = 1,2,3$. The possible four-field couplings are
\begin{eqnarray}
\nonumber 	&& 	A_{\theta_{a}}X_{(\theta_{1}, \theta_{2},\theta_{3})} A_{\theta_{b}} X^{*}_{(\theta_{1}, \theta_{2}, \theta_{3})}, \;\;\;\;\; \Phi_{\theta_{a}}X_{(\theta_{1}, \theta_{2},\theta_{3})} \Phi_{\theta_{b}} X^{*}_{(\theta_{1}, \theta_{2}, \theta_{3})}, \\  && A_{\theta_{a}}B_{(\theta_{1}, \theta_{2},\theta_{3})} A_{\theta_{b}} B^{*}_{(\theta_{1}, \theta_{2}, \theta_{3})}, \;\;\;\;\; \Phi_{\theta_{a}}B_{(\theta_{1}, \theta_{2},\theta_{3})} \Phi_{\theta_{b}} B^{*}_{(\theta_{1}, \theta_{2}, \theta_{3})} ,  \  
\end{eqnarray}
with $a , b= 1,2,3$. Based on the above couplings, the nonabelian generalization of $H_{\mu \nu\lambda}$ can then be defined as  
\begin{equation}
	H_{ijk} = dB_{ijk}+A^{l}_{i} \wedge B_{ljk} +A^{m}_{j} \wedge B_{imk} +A^{n}_{k} \wedge B_{ijn}, 
\end{equation}
with $H \sim H_{(\theta_{1},\theta_{2},\theta_{3})\mu \nu\lambda}$, $A^{l}_{i} \sim A^{l}_{(\theta_{1})\mu i}$, $A^{m}_{j} \sim A^{m}_{(\theta_{2})\mu j}$, $A^{n}_{k} \sim A^{n}_{(\theta_{3})\mu k}$, $B \sim B_{(\theta_{1},\theta_{2},\theta_{3})\mu \nu}$.

Fermions may get mass through the Yukawa coupling $\Phi_{\theta_{a}} \bar{\Psi}_{(\theta_{1}, \theta_{2}, \theta_{3})} \Psi_{(\theta_{1}, \theta_{2},\theta_{3} )}$. In order to compare with the 3-string junctions in $4d$ SYM theory, we will use $(\vec{r}, \vec{s}, \vec{t})$ instead of $(\theta_{1}, \theta_{2}, \theta_{3})$. Consider $\Psi_{(\vec{r}, \vec{s}, \vec{t})ijk}$ and $\Phi^{I}_{(\vec{u};lm)\mu}$. The vacuum expectation value of $X^{I}$ is $\bar{X}^{I}_{lm}= v^{I}_{m} \delta_{lm}$, then the induced vacuum expectation value for $\Phi^{I}_{\mu}$ is $\bar{\Phi}^{I}_{(\vec{u};lm)\mu}= \tilde{u}_{\mu} v^{I}_{m} \delta_{lm}$. Similar with the equation for fermions in \cite{3},
\begin{equation}
	\Gamma^{\mu} D_{\mu} \Psi_{A} + X^{I}_{C}C^{\nu}_{B}\Gamma_{\nu}\Gamma^{I} \Psi_{D} f^{CDB}_{\;\;\;\; \;\;\;\;A}=0,  
\end{equation}
we may have 
\begin{eqnarray}\label{asaszx}
\nonumber 	&& 	i \Gamma^{0}\Gamma^{\mu}\Gamma_{I}[ \bar{\Phi}^{I}_{(\vec{r};il)\mu}\Psi_{(\vec{r}, \vec{s}, \vec{t})ljk}+ \bar{\Phi}^{I}_{(\vec{s};jl)\mu}\Psi_{(\vec{r}, \vec{s}, \vec{t})ilk}+	 \bar{\Phi}^{I}_{(\vec{t};kl)\mu}\Psi_{(\vec{r}, \vec{s}, \vec{t})ijl}] \\ \nonumber &= & i \Gamma^{0}\Gamma^{\mu}\Gamma_{I} [\tilde{r}_{\mu} v^{I}_{i} \delta_{il} \Psi_{(\vec{r}, \vec{s}, \vec{t})ljk}+ \tilde{s}_{\mu} v^{I}_{j} \delta_{jl} \Psi_{(\vec{r}, \vec{s}, \vec{t})ilk}+\tilde{t}_{\mu}v^{I}_{k} \delta_{kl}	\Psi_{(\vec{r}, \vec{s}, \vec{t})ijl} ]\\  &= &  i \Gamma^{0}\Gamma^{\mu}\Gamma_{I}( \tilde{r}_{\mu} v^{I}_{i}  + \tilde{s}_{\mu} v^{I}_{j}  + \tilde{t}_{\mu}v^{I}_{k} )	\Psi_{(\vec{r}, \vec{s}, \vec{t})ijk} = M \Psi_{(\vec{r}, \vec{s}, \vec{t})ijk},  \  
\end{eqnarray}
where in the last step, we assume $\Psi_{lmn}=0$ for $l,m,n \neq i,j,k$ so that $\Psi$ is a generator with the index $[i,j,k]$. 
\begin{equation}\label{tot}
	M= i \Gamma^{0}\Gamma^{\mu}\Gamma_{I} (\tilde{r}_{\mu}v_{ij}^{I}-\tilde{t}_{\mu}v_{jk}^{I}) =  i \Gamma^{0}\Gamma^{\mu}\Gamma_{I} (\tilde{s}_{\mu}v_{jk}^{I}-\tilde{r}_{\mu}v_{ki}^{I}) = i \Gamma^{0}\Gamma^{\mu}\Gamma_{I} (\tilde{t}_{\mu}v_{ki}^{I}-\tilde{s}_{\mu}v_{ij}^{I}). 
\end{equation}
The $[i,j,k]$ $(\vec{r}, \vec{s}, \vec{t})$ string is the bound state of the $[i,j]$ $(\vec{r})$ and $[j,k]$ $(-\vec{t})$ strings or the $[j,k]$ $(\vec{s})$ and $[k,i]$ $(- \vec{r})$ strings or the $[k,i]$ $(\vec{t})$ and $[i,j]$ $(-\vec{s})$ strings. In (\ref{tot}), the mass of the bound state is expressed in terms of the component strings. $\tilde{r}_{4}=2 \pi r_{5}R_{4}$, $\tilde{r}_{5}= 2 \pi r_{4}R_{5}$, $\tilde{r}_{\mu}=0$, for $\mu = 0,1,2,3$, so 
\begin{equation}
	i \Gamma^{0}\Gamma^{\mu}\Gamma_{I} \tilde{r}_{\mu} v^{I}_{i} = i \Gamma^{0}\Gamma^{4}\Gamma_{I} 2 \pi r_{5}R_{4} v^{I}_{i}+i \Gamma^{0}\Gamma^{5}\Gamma_{I} 2 \pi r_{4}R_{5}v^{I}_{i}, 
\end{equation}
and similarly for $\tilde{s}_{\mu}$ and $\tilde{t}_{\mu}$. As a result,   
\begin{equation}
	M= i \Gamma^{0}\Gamma^{4}\Gamma_{I} (\tilde{r}_{4} v^{I}_{i}  + \tilde{s}_{4} v^{I}_{j}  +	 \tilde{t}_{4}v^{I}_{k} ) + i \Gamma^{0}\Gamma^{5}\Gamma_{I} (\tilde{r}_{5} v^{I}_{i}  + \tilde{s}_{5} v^{I}_{j}  +	 \tilde{t}_{5}v^{I}_{k})= i \Gamma^{0}\Gamma^{4}\Gamma_{I}  Q^{I}_{M} + i \Gamma^{0}\Gamma^{5}\Gamma_{I} Q^{I}_{E}.    
\end{equation}
where $Q^{I}_{E}$ and $Q^{I}_{M}$ are the electric and the magnetic charge vectors in $4d$ SYM theory.
\begin{equation}
M^{2} = |\vec{Q}_{E}|^{2} + |\vec{Q}_{M}|^{2}+\Gamma_{I}\Gamma_{J} \Gamma^{4}\Gamma^{5}(Q^{I}_{M}Q^{J}_{E}-Q^{J}_{M}Q^{I}_{E}). 
\end{equation}
The third term is a matrix, nevertheless,
\begin{equation}
\sqrt{[\Gamma_{I}\Gamma_{J}\Gamma^{4}\Gamma^{5}(Q^{I}_{M}Q^{J}_{E}-Q^{J}_{M}Q^{I}_{E})]^{2} }=  2|\vec{Q}_{E} \times \vec{Q}_{M} |
\end{equation}
The above result can be compared with the mass of the 3-string junctions in $4d$ SYM theory, which is
\begin{equation}
	Z^{2}_{+} = |\vec{Q}_{E}|^{2} + |\vec{Q}_{M}|^{2}+ 2|\vec{Q}_{E} \times \vec{Q}_{M} |.  
\end{equation}

The mass term together with $i \Psi^{+}\Gamma_{\mu}\partial^{\mu}\Psi$ gives the energy 
\begin{equation}
	E = \Gamma_{0}\Gamma_{\mu}p^{\mu}+ i \Gamma^{0}\Gamma^{4}\Gamma_{I}  Q^{I}_{M} + i \Gamma^{0}\Gamma^{5}\Gamma_{I} Q^{I}_{E},
\end{equation}
where $\mu = 1,2,3,4,5$, $\Gamma_{\mu}^{+} = -\Gamma_{\mu}$. 
\begin{equation}\label{bb}
	E^{2} = |\vec{p}|^{2}+ |\vec{Q}_{E}|^{2} + |\vec{Q}_{M}|^{2}+ \Gamma_{I}\Gamma_{J} \Gamma^{4}\Gamma^{5}(Q^{I}_{M}Q^{J}_{E}-Q^{J}_{M}Q^{I}_{E}) + 2 i \Gamma_{I} (Q^{I}_{M}p^{4}+Q^{I}_{E}p^{5}). 
\end{equation}
(\ref{bb}) can be rewritten as
\begin{eqnarray}\label{oo}
\nonumber E^{2} 	&=& 	p_{a}p_{a} +(Q^{45}_{E}Q^{45}_{E}+ Q^{45}_{M}Q^{45}_{M})+ (Q_{E}^{I}Q_{E}^{I}+ Q_{M}^{I}Q_{M}^{I} ) \\  &+&  \Gamma_{I}\Gamma_{J} \Gamma^{4}\Gamma^{5}(Q^{I}_{M}Q^{J}_{E}-Q^{J}_{M}Q^{I}_{E}) + 2 i \Gamma_{I} (Q^{45}_{E}Q^{I}_{M}- Q^{I}_{E}Q^{45}_{M} ),   \  
\end{eqnarray}
where $a=1,2,3$. $Q^{45}_{E} = p^{4}$, $Q^{45}_{M} = -p^{5}$. $p_{4}$ and $p_{5}$ enter the energy formula as another charge vector $Q^{45}_{E}$ and $Q^{45}_{M}$. $p^{1}$, $p^{2}$ and $p^{3}$ appear as the normal transverse momentum. In $4d$ SYM theory, with $v^{45}_{i}$ and $v^{I}_{i}$ turned on, the energy of the 3-string junction carrying the transverse momentum $(p^{1},p^{2},p^{3})$ is consistent with (\ref{oo}). The above result can be compared with the $6d$ SYM theory, for which, 
\begin{equation}
	E = \Gamma_{0} \Gamma_{\mu}p^{\mu}+ \Gamma_{0}\Gamma_{I}v^{I}_{ij}, 
\end{equation}
so 
\begin{equation}
	E^{2} = p_{\mu}p_{\mu}+ v^{I}_{ij}v^{I}_{ij}, 
\end{equation}
which is the energy of a particle with the rest mass $\sqrt{v^{I}_{ij}v^{I}_{ij}}$ carrying the $5d$ momentum $p^{\mu}$. Now, we have different Dirac operator, giving rise to a dispersion relation different from the standard $\sqrt{m^{2}+p^{2}}$ type. $\sqrt{m^{2}+p^{2}}$ is the dispersion relation for a Lorentz invariant theory. The 3-string junctions breaks the $SO(5,1)$ symmetry into $SO(3,1)$.

$A_{N-1}$ $6d$ $(2,0)$ theory compactified on a Riemann surface $\Sigma_{g}$ with the genus $g>1$ could be decomposed into the $T_{N}$ part and the $I_{N}$ part. Each $T_{N}$ part has the $SU(N)^{3}$ symmetry, while each $I_{N}$ part gives a $SU(N)$ gauge group \cite{D1,D2}. Still, there are two sets of fields with the index $[i,j,k]$ and $[i,j]$ which may couple with each other, quite like what we have discussed above. This is not accidental. The 3-string junction on $D3$, when lifted to M theory, corresponds to $M2$ with three boundaries, which may be denoted by $M2(C_{1},C_{2},C_{3})$, with $C_{1} \sim \vec{r}$, $C_{2} \sim \vec{s}$, $C_{3} \sim \vec{t}$ \cite{qwert}. $M2$ with two boundaries is $M2(C)$. $M2(C_{1},C_{2},C_{3})$ and $M2(C)$ may couple at the boundary as long as $C=C_{1}$, or $C=C_{2}$, or $C=C_{3}$, while the product is still $M2(C_{1},C_{2},C_{3})$. Likewise, the $T_{N}$ part of the Riemann surface offers the nontrivial 1-cycles $[C_{1}]$, $[C_{2}]$, $[C_{3}]$ for $M2$ to end. $[C_{1}]+[C_{2}]+[C_{3}]=0$. Each $M2([C_{1}],[C_{2}],[C_{3}])$ can only couple with the adjacent $M2([C_{1}])$, $M2([C_{2}])$, and $M2([C_{3}])$.

The $\Sigma_{g}$ theory has $3(g-1)$ $SU(N)$ gauge groups associated with the $3(g-1)$ 1-cycles. Similarly, the $6d$ $(2,0)$ theory may contain a series of $SU(N)$ groups associated with the selfdual strings labeled by $\theta$. A different way to decompose $\Sigma_{g}$ will give a different set of $3(g-1)$ 1-cycles, for which, the corresponding $4d$ theory is S-dual to the previous one. Likewise, selfdual strings parallel to a different plane may give a different $6d$ theory which is U-dual to the original one. The situation is different for the $6d$ SYM theory, in which, there is only one gauge group. Even if the $6d$ SYM theory is compactified on a Riemann surface with $g>1$, there is still only one gauge group, while the resulting $4d$ theory is unique without the dual version. The reason is that the basic excitations on $M5$ is line-like, while the basic excitations on $D5$ is point-like. $M5$ compactified on $\Sigma_{g}$ has the richer structure than $D5$.

\subsection{The situation on coincident $M5$ branes}

In above discussion, we didn't pay too much attention to the condition (\ref{wi}). For the given $v^{I}_{i}$ and $(\vec{r}, \vec{s}, \vec{t})$, the allowed $(P_{4},P_{5})$ are not arbitrary. Especially, when $v^{I}_{ij}=0$, $\forall \: i,j$, no $(P_{4},P_{5})$ can satisfy (\ref{wi}). Nevertheless, when $P_{4}=0$ or $P_{5}=0$, the equality can be saturated, while the bound states are at the threshold or just decay. If they do not decay, then (\ref{wittt}) and (\ref{wii}) should be replace by
\begin{equation}\label{asb}
	X_{(\theta_{1},\theta_{2},\theta_{3})ijk}(x_{m}, \sin (\theta-\theta_{1}) \rho) =\sin (\theta_{2}-\theta_{1})\int dp_{\theta_{1}} \;
e^{ip_{\theta_{1}}\rho \sin (\theta-\theta_{1})} \phi_{(\theta_{1},\theta_{2},\theta_{3}; p_{\theta_{1}},0)ijk}(x_{m})
\end{equation}
and
\begin{equation}\label{assb}
X_{(\vec{r}, \vec{s}, \vec{t})ijk}(x_{m}, r_{4}x_{4}R_{5}-r_{5}x_{5}R_{4})=\sum_{k} \;
e^{ik(\frac{r_{4}x_{4}}{R_{4}}- \frac{r_{5}x_{5}}{R_{5}})}\phi_{(\vec{r}, \vec{s}, \vec{t}; k,0)ijk}(x_{m}). 	
\end{equation}
(\ref{asb}) and (\ref{assb}) are translation invariant along the $\theta_{1}$ direction and the $(r_{5}R_{4},r_{4}R_{5})$ direction respectively. They are the zero mode of the original $6d$ field (\ref{wittt}) and (\ref{wii}) along the $\theta_{1}$ and the $(r_{5}R_{4},r_{4}R_{5})$ directions. $\phi_{(\vec{r}, \vec{s}, \vec{t}; k,0)ijk}(x_{m})$ is the $4d$ $1/4$ BPS field in $V_{4} \otimes V_{in}$ multiplet. Summing over all possible $k$ will give a $5d$ field in $V_{5} \otimes V_{in}$ multiplet. (\ref{asb}) and (\ref{assb}) are in the $V_{5} \otimes V_{in}$ multiplet. They are actually the bound state of the $[i,j]$ $(r_{4},r_{5})$ string with momentum $(kr_{4}/R_{4},-kr_{5}/R_{5})$ and the $[j,k]$ $(-t_{4},-t_{5})$ string with momentum $(0,0)$. As is mentioned before, the $4d$ $(0,0)$ mode of the $6d$ field is unique, so $(-t_{4},-t_{5})$ should be fixed, while the field in (\ref{asb}) and (\ref{assb}) could simply be denoted by $X_{(\theta_{1})ijk}(x_{m}, \sin (\theta-\theta_{1}) \rho)$ and $X_{(\vec{r})ijk}(x_{m}, r_{4}x_{4}R_{5}-r_{5}x_{5}R_{4})$. Although $X_{(\theta_{1})ijk}$ or $X_{(\vec{r})ijk}$ is a $5d$ field, with all $\theta_{1}$ or $(r_{4},r_{5})$ included, the $6d$ field can be recovered again.
\begin{equation}
	\phi_{(\vec{r}; k)ijk}(x_{m})\phi'_{(\vec{r}; g)li}(x_{m}) \sim \phi''_{(\vec{r}; k+g)ljk}(x_{m}). 
\end{equation}
$X_{(\vec{r})ijk}$ or $X_{(\theta_{1})ijk}$ can only couple with $X_{(\vec{r})li}$ or $X_{(\theta_{1})li}$. Both of them are translation invariant along the same direction, so the coupling is still $5$ dimensional. Now, we have $f_{ij}(\theta, x_{\mu})$ together with $g_{ijk}(\theta, x_{\mu})$ subject to the constraints $C_{\mu}(\theta)\partial^{\mu}f_{ij}(\theta, x_{\mu}) = 0$ and $C_{\mu}(\theta)\partial^{\mu}g_{ijk}(\theta, x_{\mu}) = 0$, giving rise to the $6d$ fields. Fields related with different $\theta$ cannot couple with each other. It must be admitted that such scenario is not quite interesting.

\section{The momentum-carrying BPS states in $5d$ SYM theory}

Until now, all of the discussions are carried out in $6d$ theory's framework, in which the KK modes are fields. The $6d$ tensor multiplet field compactified on $x_{5}$ gives the $5d$ massless vector multiplet field and a tower of $5d$ massive tensor multiplet fields. As the zero mode, the $5d$ SYM field must have the vanishing Pontryagin number. However, the generic configurations of the $5d$ SYM theory on $R^{4}$ can carry the arbitrary Pontryagin number $k$, while the quantization of the configurations with the nonzero $k$ gives the $5d$ massive tensor multiplets. So the full $5d$ SYM theory contains the complete KK modes and may give another definition of the $6d$ $(2,0)$ theory \cite{41, 42}\footnote{See \cite{1112} for a further evidence on the finiteness of the $5d$ SYM theory.}.

The field configurations in SYM theories are classified by the the boundary topology. For $5d$ SYM theory, the boundary configurations are characterized by $\Pi_{3}(SU(N))\cong \textbf{Z}$ with $k \in \textbf{Z}$ the winding number. Configurations with the same $k$ could be continuously deformed into each other. Especially, when $k=0$, fields could be continuously deformed to zero. The sector with the given $k$ corresponds to the KK mode with $P_{5} = k/R_{5}$. The energy is bounded by 
\begin{equation}
	E \geq |k|/R_{5}. 
\end{equation}
The equality holds for configurations representing the localized $k/R_{5}$ mode which have the zero average momentum in $1234$ space. The path integral covers all configurations, so the complete $5d$ SYM theory is intrinsically a $6d$ theory. Since the configuration only carries the chargeless $P_{5}$ momentum, there might be some kind of confinement happen.

In this section, we will discuss the generic BPS states in $5d$ SYM theory, which are in one-to-one correspondence with the previous mentioned selfdual strings and the string junctions. We will also show that the selfdual strings carrying the longitudinal momentum have the $N^{3}$ scaling.

\subsection{BPS states in $5d$ SYM theory}

The field content of the $5d$ ${\cal N}=2$ $U(N)$ SYM theory consists of a vector $A_{\mu}$ with $\mu = 0, 1, 2, 3, 4$, five scalars $X^{I}$ with $I = 6, 7, 8, 9, 10$ and fermions $\Psi$. $x_{5}$ is the extra dimension associated with M-theory. The action is 
\begin{eqnarray}
\nonumber S &=& -\frac{1}{g^2_{YM}}\int d^5x\; {\rm tr} \Big(\frac{1}{4}F_{\mu\nu} F^{\mu\nu} + \frac{1}{2}D_\mu X^I D^\mu X^I -\frac{i}{2}\bar\Psi\Gamma^\mu D_\mu \Psi \\  &&\hskip3cm+ \frac{1
}{2}\bar\Psi \Gamma^5\Gamma^I[X^I,\Psi]- \frac{1}{4}\sum_{I,J}[X^I,X^J]^2\Big)\ ,
\end{eqnarray}
where $D_\mu X^I = \partial_\mu X^I - i[A_\mu,X^I]$, $F_{\mu\nu} = \partial_\mu A_\nu - \partial_\nu A_\mu - i [A_\mu,A_\nu]$. For time-independent bosonic solutions with a single non-vanishing scalar field $X^{6}$, the associated energy is
\begin{equation}
	E  = \frac{1}{g^2_{YM}} \int d^4 x \;{\rm tr} \Big[ \frac{1}{4}F_{ij}F_{ij} + \frac{1}{2}F_{0i}F_{0i}+\frac{1}{2} D_iX^6D_iX^6 \Big],  
\end{equation}
where $i = 1, 2, 3, 4$. For an arbitrary vector $C_{i}$ with $|C| = 1$, $E$ could be rewritten as 
\begin{eqnarray}
\nonumber E &=& \frac{1}{g^2_{YM}} \int d^4 x \;{\rm tr} \Big[ \frac{1}{2}(F_{0i} - \sin\theta C_{k}F_{ik} + \cos\theta D_{i}X^{6} )^{2}  \\ \nonumber &&\hskip2.5cm + \frac{1}{2}(\frac{1}{2}C_{k}\varepsilon_{ilmk}F_{lm} \pm \cos\theta C_{k}F_{ik} \pm \sin\theta  D_{i}X^{6} )^{2} \\ \nonumber &&\hskip2.5cm + \sin\theta \:(F_{0i}C_{k}F_{ik}\mp \frac{1}{2}C_{k}\varepsilon_{ilmk}F_{lm}D_{i}X^{6}) \\  &&\hskip2.5cm +\cos\theta \: (\mp \frac{1}{8} \varepsilon_{iklm}F_{ik}F_{lm}  - F_{0i}D_{i}X^{6})\Big]\ . 
\end{eqnarray}
Note that 
\begin{eqnarray}\label{as}
	\nonumber && P_{k} = -\frac{1}{g^2_{YM}} \int d^4 x \;{\rm tr} (F_{0i}F_{ik}),\;\;\;  Q_{Mk}= Z^{6}_{k} = - \frac{1}{2g^2_{YM}} \int d^4 x \;{\rm tr}(\varepsilon_{iklm}F_{lm}D_{i}X^{6}),\\  && 
 P_{5} = -\frac{1}{8g^2_{YM}} \int d^4 x \;{\rm tr} (\varepsilon_{iklm}F_{lm}F_{ik}),\;\;\;  Q_{E}= Z^{6}_{5} =  \frac{1}{g^2_{YM}} \int d^4 x \;{\rm tr}(F_{0i}D_{i}X^{6}),  
\end{eqnarray}
So
\begin{equation} \label{3}
E  \geq  \sin\theta C_{k}(-P_{k}\pm Q_{Mk})+ \cos\theta (\pm P_{5}-Q_{E}) \geq  Max (Z_{+},Z_{-}), 	
\end{equation}
where 
\begin{equation}
	Z_{\pm} = \Big[ (C_{k}P_{k}\pm C_{k}Q_{Mk})^{2}+ (P_{5} \pm Q_{E})^{2}\Big]^{\frac{1}{2}}. 
\end{equation}
If $Z_{+} \geq Z_{-}$, $E = Z_{+}$ for
\begin{equation}\label{1}
	F_{0i} = \sin\theta C_{k}F_{ik} - \cos\theta D_{i}X^{6},
\end{equation}
\begin{equation}\label{2}
	\frac{1}{2}C_{k}\varepsilon_{ilmk}F_{lm} = \cos\theta C_{k}F_{ik} + \sin\theta  D_{i}X^{6}. 
\end{equation}
If $Z_{+} \leq Z_{-}$, $E = Z_{-}$ for
\begin{equation}\label{12}
	F_{0i} = \sin\theta C_{k}F_{ik} - \cos\theta D_{i}X^{6},
\end{equation}
\begin{equation}\label{21}
	\frac{1}{2}C_{k}\varepsilon_{ilmk}F_{lm} = - \cos\theta C_{k}F_{ik} - \sin\theta  D_{i}X^{6}. 
\end{equation}
In both cases, 
\begin{equation}
	 E = \frac{1}{g^2_{YM}} \int d^4 x \;{\rm tr} \Big[(C_{k}F_{ik})^{2}+ (D_{i}X^{6})^{2}\Big]. 
\end{equation}
Moreover, if $\theta \neq 0$, from (\ref{1}-\ref{21}), we also have $C_{i}D_{i}X^{6} = C_{i}F_{0i} = 0$. For simplicity, in the following, we will only consider the case with $Z_{+} \geq Z_{-}$. The situation with $Z_{+}\leq Z_{-}$ is similar.

Without loss of generity, let $C_{k} = \delta^{4}_{k}$, then (\ref{1}) and (\ref{2}) become
\begin{equation}\label{5}
F_{0i} = \sin\theta F_{i4} - \cos\theta D_{i}X^{6},
\end{equation}
\begin{equation}\label{6}
	\frac{1}{2}\varepsilon_{ilm4}F_{lm} = \cos\theta F_{i4} + \sin\theta  D_{i}X^{6}.
\end{equation}
\begin{equation}\label{KN1}
	E = \Big[ (P_{4}+ Q_{M4})^{2}+ (P_{5} + Q_{E})^{2}\Big]^{\frac{1}{2}}. 
\end{equation}
For $\theta \neq 0$, $F_{04} = D_{4}X^{6} = 0$. When $\theta = 0$, (\ref{5})-(\ref{KN1}) reduce to 
\begin{equation}\label{KN}
	F_{0i}=-D_{i}X^{6}, \;\;\;\;\;\frac{1}{2}\varepsilon_{ilm4}F_{lm} = F_{i4}. 
\end{equation}
\begin{equation}
	E = |P_{5}+Q_{E}|. 
\end{equation}
These are the equations for the dyonic instantons discussed in \cite{42}. $F_{04} = D_{4}X^{6} = 0$ is not necessary. If is imposed, the original $SO(4)$ symmetry will be broken to $SO(3)$. $P_{4}\neq 0$, $Q_{M4} \neq 0$, but $P_{4}+Q_{M4} = 0$. When $\theta = \pi /2$,
\begin{equation}
	F_{0i} = F_{i4} , \;\;\;\;\;\frac{1}{2}\varepsilon_{ilm4}F_{lm}  = D_{i}X^{6}.
\end{equation}
 \begin{equation}
	E = |P_{4}+Q_{M4}|. 
\end{equation}
The solution describes the monopole string extending along the $x_{4}$ direction, carrying momentum $P_{4}$. $P_{5}\neq 0$, $Q_{E} \neq 0$, but $P_{5}+Q_{E} = 0$.

For the time-independent bosonic solutions with $C_{k}=\delta^{4}_{k}$, the supersymmetry transformation becomes 
\begin{eqnarray}\label{4}
\nonumber 	\delta_\epsilon\Psi  &=&
\frac{1}{2}F_{\mu\nu}\Gamma^{\mu\nu}\Gamma_{5}\epsilon + D_\mu X^6\Gamma^\mu\Gamma^6\epsilon  \\ \nonumber &= & D_{a}X^{6} \Gamma_{a}(\Gamma^{6}+ \cos\theta \Gamma_{05}+ \sin\theta \Gamma_{123}\Gamma_{5})\epsilon  \\   && + F_{a4} \Gamma_{a} (\Gamma_{45}- \sin\theta \Gamma_{05} + \cos\theta \Gamma_{123}\Gamma_{5} )\epsilon, \  
\end{eqnarray}
where $a = 1, 2, 3$. $D_{4}X^{6} = F_{04} = 0$ is imposed. $\delta_\epsilon\Psi = 0$, $\epsilon$ should satisfy 
\begin{equation}
	(1+ \cos\theta \Gamma_{056}- \sin\theta \Gamma_{046})\epsilon= 0, 
\end{equation}
\begin{equation}
	(1+ \sin\theta \Gamma_{04} - \cos\theta \Gamma_{05})\epsilon = 0,  
\end{equation}
in which $\Gamma_{012345}\epsilon = \epsilon$ is used. The solution is $1/4$ BPS. For $\theta = 0$, we have $\epsilon = - \Gamma_{056} \epsilon = \Gamma_{05}\epsilon$, which are the supersymmetries preserved by dyonic instantons \cite{42}. For $\theta = \pi/2$, $\epsilon =  \Gamma_{046} \epsilon = - \Gamma_{04}\epsilon$, which are the supersymmetries preserved by the monopole strings extending along $x_{4}$ carrying momentum $P_{4}$. If $D_{a}X^{6}$ and $F_{a4}$ are not independent, for example, $D_{a}X^{6} = \sin\theta D_{a} \Phi$ and $F_{a4} = \cos\theta D_{a} \Phi$ as that in \cite{42}, (\ref{4}) will reduce to 
\begin{eqnarray}
\nonumber 	\delta_\epsilon\Psi &=& D_{a} \Phi \Gamma_{a}(\sin\theta  \Gamma^{6}+ \cos\theta \Gamma_{45} - \Gamma_{04})\epsilon   \\  &= & D_{a} \Phi \Gamma_{a}\Gamma_{04}( \sin\theta \Gamma_{04}  \Gamma^{6}+ \cos\theta \Gamma_{05}-1)\epsilon  = 0.   
\end{eqnarray}
The solution becomes $1/2$ BPS. Moreover, for this state, $F_{0i} = 0$, so $P_{k} = Q_{E} = 0$, $E = \sqrt{Q^{2}_{M4}+ P^{2}_{5}}$. It may describe the monopole string extending along $x_{4}$ carrying the uniformly distributed $D0$ charge. Conversely, if $F_{a4} = \sin\theta D_{a} \Phi$, $D_{a}X^{6} = - \cos\theta D_{a} \Phi$, $F_{ab} = 0$, $E = \sqrt{Q^{2}_{E}+ P^{2}_{4}}$. The solution describes the $F1$ string carrying $P_{4}$ momentum, which is also $1/2$ BPS.

Another special kind of $1/2$ BPS states have $F_{i4} = 0$ or $X^{6} = 0$. When $X^{6} = \theta = 0$, we get the instanton equation 
\begin{equation}
	F_{0i} = 0, \;\;\;\;\;\frac{1}{2}\varepsilon_{ilm4}F_{lm} = F_{i4}, 
\end{equation}
the solution of which describes the $D0$ branes revolved in $D4$ branes. $E = |P_{5}|$. The quantization of the instanton state gives the $5d$ massive $(2,0)$ tensor multiplet $T_{5}$ without charge. When $\theta \neq 0$, 
\begin{equation}
	F_{0i} = \sin\theta F_{i4}, \;\;\;\;\;\frac{1}{2}\varepsilon_{ilm4}F_{lm} = \cos\theta F_{i4}. 
\end{equation}
The $SO(4)$ symmetry is broken to $SO(3)$. Therefore, we may look for solutions which are translation invariant along $x_{4}$. $E = \sqrt{P^{2}_{4}+P^{2}_{5}}$. The solution describes the $D0$ branes localized in $R^{3}$ carrying momentum $P_{4}$, which, in $D3$ picture, is the $(p,q)$ strings winding $x'^{4}$. The quantization gives the $4d$ massive vector multiplet $V_{4}$ that is also the KK mode of the $5d$ massive tensor multiplet $T_{5}$. The original four position moduli of the instantons become the three position moduli plus one momentum moduli. $\tan \theta = P_{4}/P_{5}$. The $(p,q)$ string can be open or closed, thus carries the $[i,j]$ charge or not, so is the corresponding $4d$ vector multiplet.

On the other hand, if $F_{i4} = \theta = 0$, the equations will be 
\begin{equation}
	F_{0i} = -D_{i}X^{6}, \;\;\;\;\;\frac{1}{2}\varepsilon_{ilm4}F_{lm} =   0, 
\end{equation}
whose solutions are $[i,j]$ $F1$ strings, the quantization of which gives the $5d$ vector multiplet $V_{5}$. $E=|Q_{E}|$. When $\theta \neq 0$, 
\begin{equation}
	F_{0i} = - \cos\theta D_{i}X^{6}, \;\;\;\;\;\frac{1}{2}\varepsilon_{ilm4}F_{lm} =  \sin\theta  D_{i}X^{6}. 
\end{equation}
The solution describes the bound state of the $[i,j]$ $F1$ and the $[i,j]$ monopole string extending along $x_{4}$, whose quantization also gives the $4d$ vector multiplet $V_{4}$. $E = \sqrt{Q^{2}_{E}+Q^{2}_{M4}}$. $\tan \theta =Q_{M4}/Q_{E}$. In this case, $\theta$ is just the previously mentioned label for the selfdual strings parallel to the $45$ plane. A reduction along $x_{5}$ is made to get the states with $P_{5}=0$. Selfdual strings extending along $x_{5}$ already have $P_{5}=0$ and is projected to a point in $5d$. The rest selfdual strings are projected to a straight line extending along $x_{4}$, which is the bound state of the $[i,j]$ $F1$ and the $[i,j]$ monopole string. $F1$ has the definite momentum $P_{4}=0$, while the monopole string carries no $D0$ charge, so the bound state is the zero mode of the $6d$ theory on $x_{4}\times x_{5}$, which should be unique, but is now degenerate.

For $1/4$ BPS state, when $\theta = 0$, we get (\ref{KN}), whose solution is the dyonic instanton, the quantization of which gives the $5d$ massive $(2,1)$ multiplet with $2^{6}$ complex states composed by $1$ spin-3/2 fermion, $13$ spin-1/2 fermions, $2$ selfdual tensors, $4$ vectors and $10$ scalars \cite{42}, which is actually the previously mentioned $T_{5} \otimes ([1/2]\oplus[0]\oplus[0])$. When $\theta \neq 0$, the equations are (\ref{5}) and (\ref{6}). The solution corresponds to the bound state of the string and the monopole string, carrying the $P_{4}$ $P_{5}$ transverse momentum respectively. The string and the monopole string carry the different charge, for example, $[i,j]$ and $[j,k]$. The quantization gives the $4d$ $V_{4} \otimes ([1/2]\oplus[0]\oplus[0])$ multiplet with $2^{6}$ real states composed by $1$ spin-3/2 fermion, $14$ spin-1/2 fermions, $6$ vectors and $14$ scalars, which is the massive KK mode of $T_{5} \otimes ([1/2]\oplus[0]\oplus[0])$. Notice that for the $F1$-$D0$ bound state, $D0$ is chargeless, so the corresponding multiplet can only carry the $[i,j]$ charge. On the other hand, for the $F1$-$D2$ bound state with the transverse momentum involved, $F1$ and $D2$ may carry the $[i,j]$ and $[j,k]$ charges, and so the corresponding multiplet may have the index $[i,j,k]$. Just as the $1/2$ BPS case, $D0$ in momentum other than position eigenstate of $x_{4}$ can carry charge.

\subsection{Selfdual string carrying the longitudinal momentum}

It is convenient to work in $D3$ picture. With $x_{4}$ compactified, under the T-duality transformation along $x_{4}$, $A_{4}\rightarrow X^{4}$. Let $F_{0a} = E_{a}$, $\frac{1}{2}\epsilon_{abc4}F_{bc} = B_{a}$, (\ref{5}) and (\ref{6}) could be rewritten as 
\begin{equation}\label{7}
E_{a} = \sin\theta D_{a}X^{4} - \cos\theta D_{a}X^{6},
\end{equation}
\begin{equation}\label{8}
B_{a}= \cos\theta D_{a}X^{4} + \sin\theta  D_{a}X^{6}, 
\end{equation}
which are the standard BPS equations for the ${\cal N}=4$ SYM theory with two scalar fields $X^{4}$ and $X^{6}$ turned on. In the language of the ${\cal N}=4$ $SU(N)$ SYM theory, 
\begin{equation}
	 \int dS_{a} \; E_{a} =  e \textbf{p} \cdot \textbf{H}, \;\;\;\; \int dS_{a} \; B_{a} = \frac{4\pi}{e} \textbf{q} \cdot \textbf{H}, 
\end{equation}
where the vectors $\textbf{p}$ and $\textbf{q}$ are the electric and the magnetic charges respectively. $\textbf{H}$ generates the Cartan subalgbra of $SU(N)$. 
\begin{equation}
	\textbf{p} \cdot \textbf{H} = \textnormal{diag} (p_{1},p_{2},\cdots ,p_{N}), \;\;\;\; \textbf{q}\cdot \textbf{H}  = \textnormal{diag} (q_{1},q_{2},\cdots ,q_{N}), 
\end{equation}
$\sum^{N}_{i=1} p_{i} = \sum^{N}_{i=1} q_{i} = 0$. Suppose $\left\langle X^{I}  \right\rangle = \textbf{v}^{I} \cdot \textbf{H}  = \textnormal{diag} (v^{I}_{1},v^{I}_{2},\cdots ,v^{I}_{N})$, $v^{6}_{1}\geq v^{6}_{2} \geq \cdots \geq v^{6}_{N}$,     
\begin{equation}\label{81}
	{\cal Q}^{4}_{E} =  \int d^3 x \; \partial_{a}{\rm tr} \Big[ X^{4} E_{a}\Big] = e \textbf{p} \cdot \textbf{v}^{4}  = e \sum^{N}_{i=1} p_{i} v^{4}_{i}\sim -P_{4}, 
\end{equation}
\begin{equation}
	{\cal Q}^{6}_{E} =  \int d^3 x \; \partial_{a}{\rm tr} \Big[ X^{6} E_{a}\Big]= e \textbf{p} \cdot \textbf{v}^{6}= e \sum^{N}_{i=1} p_{i} v^{6}_{i} \sim Q_{E}, 
\end{equation}
\begin{equation}
	{\cal Q}^{4}_{M} =  \int d^3 x \; \partial_{a}{\rm tr} \Big[ X^{4} B_{a}\Big]= \frac{4\pi}{e} \textbf{q} \cdot \textbf{v}^{4}= \frac{4\pi}{e} \sum^{N}_{i=1} q_{i} v^{4}_{i}\sim -P_{5}, 
\end{equation}
\begin{equation}\label{18}
	{\cal Q}^{6}_{M} =  \int d^3 x \; \partial_{a}{\rm tr} \Big[ X^{6} B_{a}\Big]= \frac{4\pi}{e}  \textbf{q} \cdot \textbf{v}^{6}= \frac{4\pi}{e} \sum^{N}_{i=1} q_{i} v^{6}_{i} \sim -Q_{M4}.  
\end{equation}
The energy becomes\footnote{For (\ref{29}) to be valid, for the given $\textbf{v}^{I}$, $\textbf{p}$ $\textbf{q}$ should be selected so that $Z_{+}\geq Z_{-}$, otherwise $	E = \sqrt{({\cal Q}^{4}_{E}-{\cal Q}^{6}_{M} )^{2}+({\cal Q}^{6}_{E}+{\cal Q}^{4}_{M} )^{2}}$.}
\begin{equation}\label{29}
	E = \sqrt{({\cal Q}^{4}_{E}+{\cal Q}^{6}_{M} )^{2}+({\cal Q}^{6}_{E}-{\cal Q}^{4}_{M} )^{2}}. 
\end{equation}
$x^{4} \sim x^{4}+ 2\pi n R$. The transverse position of the $i_{th}$ $D3$ brane in $4$-$6$ plane could be denoted by $(v^{4}_{i}, v^{6}_{i})$, where $v_{i}^{I} \in (-\infty, +\infty)$.

The generic $1/2$ BPS state is the $(p,q)$ string connecting the $i$ $j$ $D3$ branes with the mass
\begin{equation}
	E = \sqrt{[ep(v^{6}_{i}-v^{6}_{j})- \frac{4\pi}{e}q(v^{4}_{i}-v^{4}_{j}) ]^{2}+[\frac{4\pi}{e}q(v^{6}_{i}-v^{6}_{j})+ ep(v^{4}_{i}-v^{4}_{j})]^{2}}. 
\end{equation}
Especially, if $e^{2}p(v^{6}_{i}-v^{6}_{j})= 4\pi q(v^{4}_{i}-v^{4}_{j})$, the state will reduce to a $[j,i]$ $D2$ brane carrying $[j,i]$ $P_{4}$ momentum, while if $4\pi q(v^{6}_{i}-v^{6}_{j})=- e^{2}p(v^{4}_{i}-v^{4}_{j})$, the state will become a $[i,j]$ string with $[j,i]$ $D0$ charge. Notice that in this case, the $P_{4}$ momentum and the $D0$ charge spread uniformly over the $D2$ branes and the strings.

The simplest $1/4$ BPS state is the 3-string junction with $i$ $j$ $k$ representing three distinct $D3$ branes with coordinates $(v^{4}_{i}, v^{6}_{i})$, $(v^{4}_{j}, v^{6}_{j})$, $(v^{4}_{k}, v^{6}_{k})$. With the charge vector $\textbf{v}_{e}=(1,0,-1)$, $\textbf{v}_{m}=(0,1,-1)$, the mass is 
\begin{equation}
	E = \sqrt{[e(v^{6}_{i}-v^{6}_{k})- \frac{4\pi}{e}(v^{4}_{j}-v^{4}_{k}) ]^{2}+[\frac{4\pi}{e}(v^{6}_{j}-v^{6}_{k})+ e(v^{4}_{i}-v^{4}_{k})]^{2}}. 
\end{equation}
The corresponding state on $D4$ is a $[i,k]$ string with $v^{4}_{k}-v^{4}_{j}$ $D0$ charge and a $[k,j]$ $D2$ brane with $v^{4}_{k}-v^{4}_{i}$ $P_{4}$ momentum. Especially, when $e^{2}(v^{6}_{i}-v^{6}_{k} )= 4 \pi (v^{4}_{j}-v^{4}_{k})$, the state reduces to the $[k,j]$ $D2$ brane carrying $[k,i]$ $P_{4}$ momentum, while when $4 \pi (v^{6}_{j}-v^{6}_{k}) =e^{2} (v^{4}_{k}-v^{4}_{i})$, the state reduces to the $[i,k]$ string carrying $[k,j]$ $D0$ charge. The $[i,k]$ string and the $[k,j]$ $D2$ brane are parallel, so the bound state does not exist, nevertheless, with suitable amount of $P_{4}$ momentum and the $D0$ charge, the bound state may form. With the given $A_{4}$, the $[i,j]$ string ($D2$ brane) can only carry the $[i,j]$ $P_{4}$ momentum ($D0$ charge). However, they can carry the $[j,k]$ or $[i,k]$ $D0$ charge ($P_{4}$ momentum), which is actually the transverse momentum of the $[j,k]$ or $[i,k]$ $D2$ brane (string). The $[i,j]$ selfdual string carrying the $[j,k]$ longitudinal momentum has the $N^{3}$ scaling.

Now, consider string webs with more external legs. For $i \geq k \geq l \geq j$, the $[i,j]$ $D2$ brane (string) $[k,l]$ string ($D2$ brane) bound states do not exist. The bound state may exist if the $[k,l]$ string ($D2$ brane) carries the appropriate $P_{4}$ momentum ($D0$ charge). In general, the charge vector can be taken as $p_{i}=1$, $p_{j}=-1$, $p_{a}=0$ for $a \neq i, j$, $q_{a}=0$ for $a > i$ or $a< j$. 
\begin{eqnarray}
\nonumber  && 
P_{4} = e (v^{4}_{j} - v^{4}_{i}), \;\;\;\; Q_{E} = e (v^{6}_{i} - v^{6}_{j}),   \\ \nonumber  &&  P_{5} = -\frac{4\pi}{e}\sum^{i}_{m=j} q_{m} v^{4}_{m} = \frac{4\pi}{e}\sum^{i}_{m=j} r_{m} (v^{4}_{m+1}-v^{4}_{m}),  \\   && Q_{M4} = -\frac{4\pi}{e}\sum^{i}_{m=j} q_{m} v^{6}_{m} = \frac{4\pi}{e}\sum^{i}_{m=j} r_{m} (v^{6}_{m+1}-v^{6}_{m}).   
\end{eqnarray}  
This is the bound state of the $[i,j]$ string and $r_{m}$ $[m+1,m]$ $D2$ branes each carrying $v^{4}_{m+1}-v^{4}_{m}$ $[m+1,m]$ $D0$ charge. Especially, if the $[i,j]$ string carries the transverse momentum $P_{4}$ so that $P_{4}+ Q_{M4} = 0$, the state will reduce to the $[i,j]$ string with $r_{m}$ $v^{4}_{m+1}-v^{4}_{m}$ $[m+1,m]$ $D0$ charge. Conversely, one may let $q_{i}=1$, $q_{j}=-1$, $q_{a}=0$ for $a \neq i, j$, $p_{a}=0$ for $a > i$ or $a< j$. 
\begin{eqnarray}
\nonumber  && P_{5} = \frac{4\pi}{e}(v^{4}_{j} - v^{4}_{i}), \;\;\;\;Q_{M4} = \frac{4\pi}{e}(v^{6}_{j} - v^{6}_{i})\\ \nonumber  && P_{4} = -e \sum^{i}_{m=j} p_{m} v^{4}_{m} =e \sum^{i}_{m=j} s_{m} (v^{4}_{m+1}-v^{4}_{m}),  \\   &&
Q_{E} = e \sum^{i}_{m=j} p_{m} v^{6}_{m} = - e \sum^{i}_{m=j} s_{m} (v^{6}_{m+1}-v^{6}_{m}).    
\end{eqnarray}  
The corresponding state is the bound state of the $[j,i]$ $D2$ brane and $s_{m}$ $[m,m+1]$ strings each carrying the $v^{4}_{m+1}-v^{4}_{m}$ $P_{4}$ momentum. When $P_{5}+ Q_{E} =0$, the state becomes $[j,i]$ $D2$ brane carrying $s_{m}$ $v^{4}_{m+1}-v^{4}_{m}$ $[m+1,m]$ $P_{4}$ momentum.

We can give a more precise description for these longitudinal-momentum-carrying states. For example, for string web in Fig.1, suppose the strings extending in $x_{4}$ $x_{6}$ directions are $(1,0)$ and $(0,1)$ strings, while the rest ones are $(1,1)$ strings, then the state could be taken as the $[i,n]$ $D2$ extending along $x_{4}\times x_{6}$, for which, the $[i,j]$ $[k,l]$ $[m,n]$ $D2$ carry the zero $P_{4}$ momentum, the $[j,k]$ $[l,m]$ $D2$ have the uniformly distributed $T_{F1}|v^{4}_{ab}|$, $T_{F1}|v^{4}_{cd}|$, $P_{4}$ momentum, while the rest $T_{F1}|v^{4}_{ja}|$, $T_{F1}|v^{4}_{bk}|$, $T_{F1}|v^{4}_{lc}|$, $T_{F1}|v^{4}_{dm}|$, $P_{4}$ momentums are localized on the $j_{th}$, $k_{th}$, $l_{th}$, $m_{th}$ $D4$ branes.   
\begin{figure}[htbp]
\begin{center}
\includegraphics[width=10cm, height=6cm]{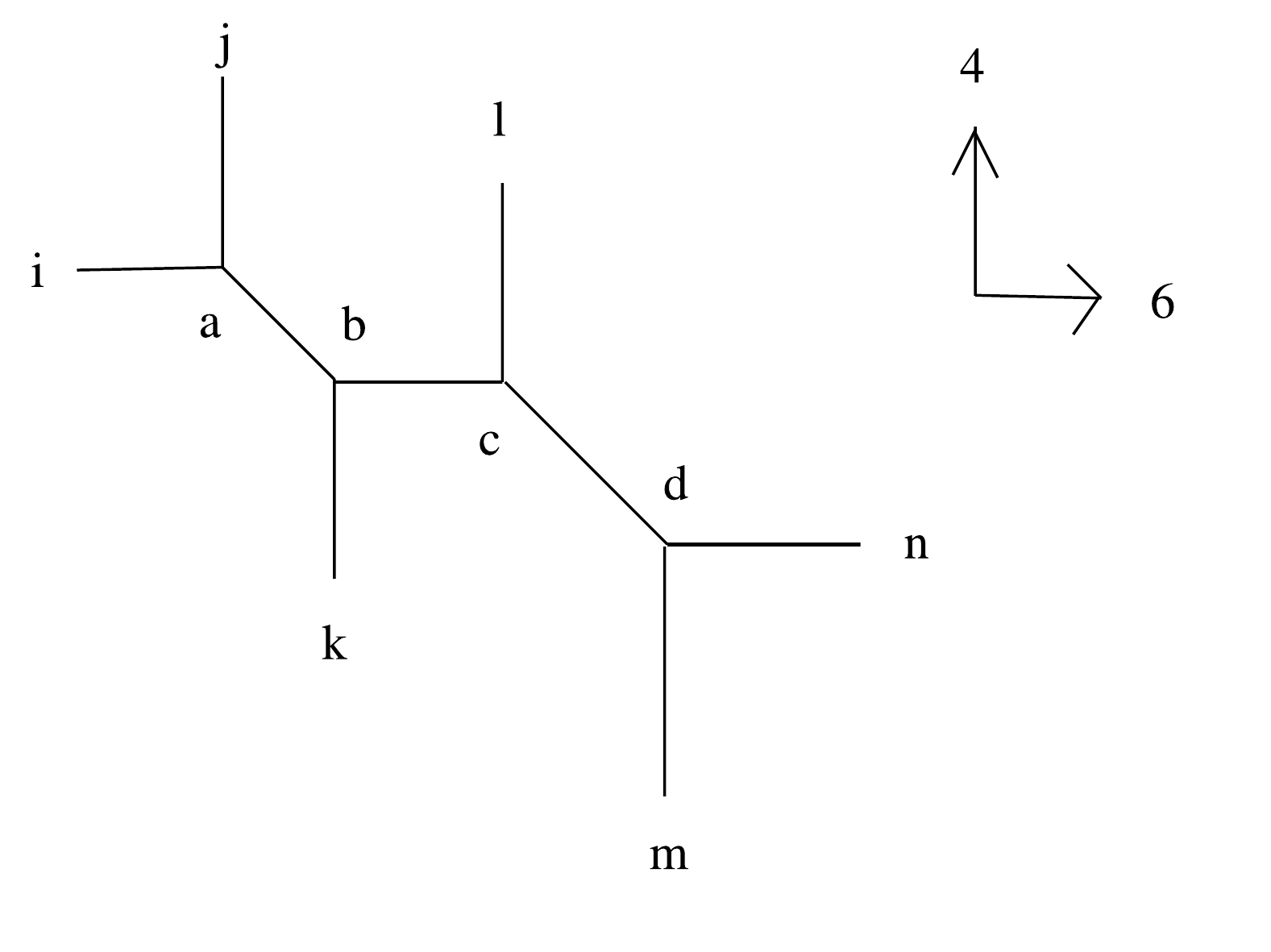}
\caption{The $ijklmn$ string web}
\label{fig:0}
\end{center}
\end{figure}

$[i,j]$ $D2$ ($F1$) is composed by $[i,i+1]$ $\cdots$ $[j-1,j]$ $D2$'s ($F1$'s). Each $[a,a+1]$ $D2$ ($F1$) must have the same transverse velocity, otherwise, the bound state cannot be formed. On the other hand, the longitudinal momentums along $x_{4}$ ($x_{5}$) on each $[a,a+1]$ $D2$ ($F1$) are independent, so the degrees of freedom on the $[i,j]$ $D2$ ($F1$) is $j-i$. Altogether, there are $N(N-1)/2$ $[i,j]$ $D2$ ($F1$), therefore, the total number of degrees of freedom is $(N^{3}-N)/6$. The $N^{3}$ scaling comes from the longitudinal momentum. Both transverse momentum and the longitudinal momentum carry the charge. The $[i,j]$ $D2$ ($F1$) can only carry the $[i,j]$ transverse momentum but the $[k,l]$ longitudinal momentum for any $i\leq k < l\leq j$. The index calculation in \cite{ct} also showed the $(N^{3}-N)/6$ degrees of freedom for the longitudinal momentum mode on open $D2$'s connecting $D4$'s.

\section {The degrees of freedom at the triple intersection of $M5$ branes}

In this section, we will consider the the triple intersecting configuration of the $M5$ branes $5 \bot 5 \bot 5$. We will discuss the possible string junctions and their relevance with the $N^{3}$ degrees of freedom at the triple interaction.

Suppose there are $N_{1}$, $N_{2}$, $N_{3}$, $M5$ branes extending in $0$ $1$ $2$ $3$ $4$ $5$ direction, $0$ $1$ $2$ $3$ $6$ $7$ direction, and $0$ $1$ $4$ $5$ $6$ $7$ direction respectively (see Fig.2). 
\begin{figure}[htbp]
\begin{center}
\includegraphics[width=10cm, height=4cm]{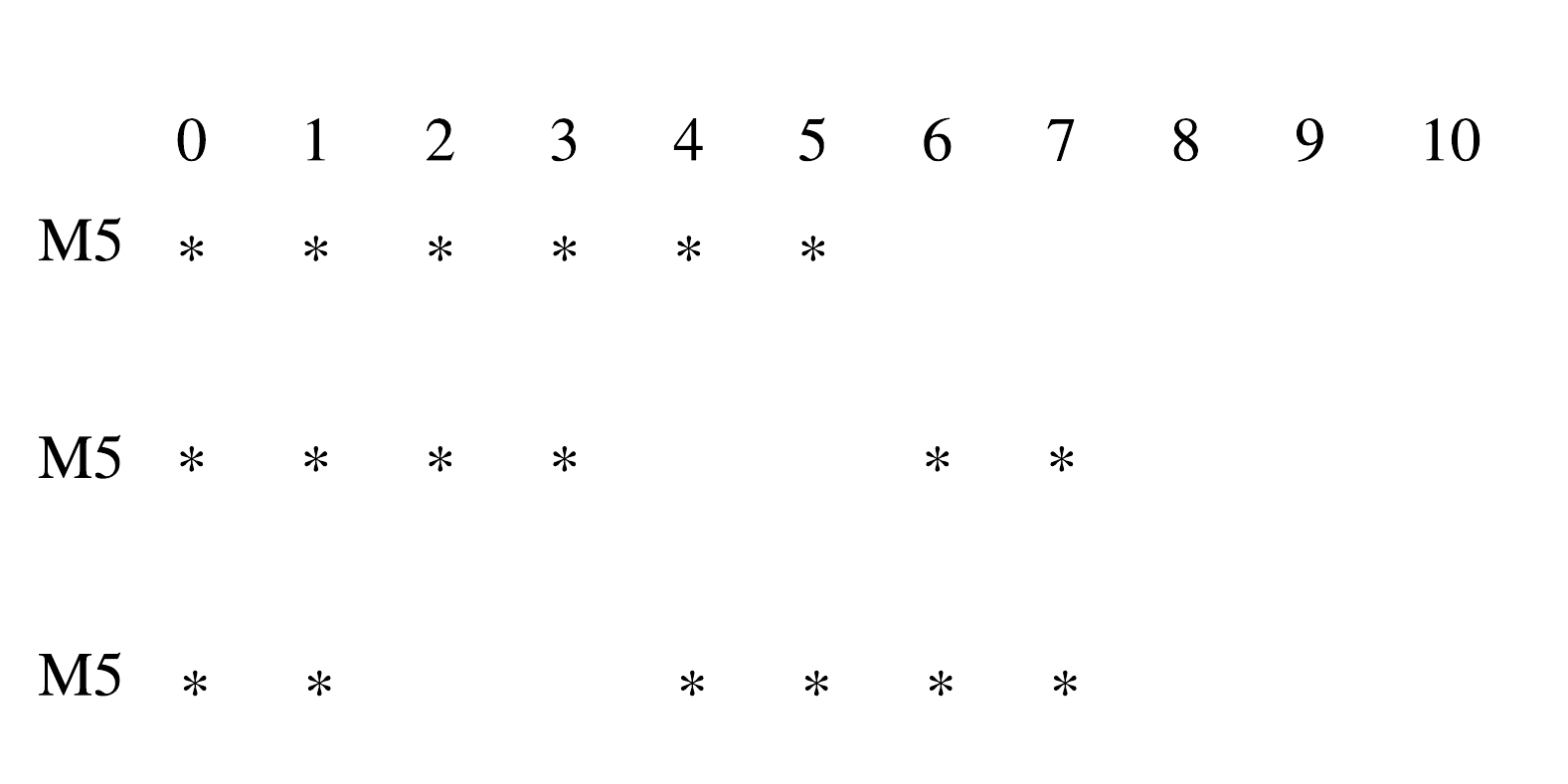}
\caption{The $M5$ $M5$ $M5$ configuration}
\label{fig:1}
\end{center}
\end{figure}
The common transverse space is $x_{8}$ $x_{9}$ $x_{10}$, while the common longitudinal spacetime is $x_{0}$ $x_{1}$ with $x_{0}$ the time direction. If $x_{8}=x_{9}=x_{10}$, the $N_{1}+N_{2}+N_{3}$ $M5$ branes will have $N_{1}N_{2}N_{3}$ triple intersections no matter whether each bunch of $M5$ branes are coincident or not. 
The black hole entropy calculation shows that there are $N_{1}N_{2}N_{3}$ degrees of freedom at the triple intersections, so each intersection will offer one degree of freedom \cite{qaz}. The situation can be compared with the $4 \bot 4$ configuration for $N_{1}$ and $N_{2}$ intersecting $D4$ branes with $N_{1}N_{2}$ $3d$ intersections. There are $U(1)\times U(1)$ massless hypermultiplets living at each intersection, producing the $N_{1}N_{2}$ entropy. So, we may expect that similarly the triple intersection will also capture some nonabelian features of $M5$.

Consider one intersection and label the three $M5$ branes by $i$, $j$, $k$. In the most generic case, $i$ $j$ $k$ $M5$ branes appear as three points $v^{I}_{i}$ $v^{I}_{j}$ $v^{I}_{k}$ in $x_{8}\times x_{9} \times x_{10}$ transverse space. Still, we want to compactify two longitudinal dimensions of $M5$ branes to simplify the problem. There are two distinct possibilities: $x_{2} \times x_{4}$ and $x_{2} \times x_{1}$. $M$ theory compactified on $x_{2}$ gives the type IIA string theory, with the $i$ $j$ $k$ $M5$'s becoming the $D4$ $D4$ $NS5$. The triple intersection of the $D4$ $D4$ $NS5$ branes still have the $N_{1}N_{2}N_{3}$ entropy, so the KK mode along $x_{2}$ can be safely dropped\footnote{The $P_{2}$ momentum may have the relevance with the $c=6$ central charge. With one $M5$ fixed, there are $4$ moduli to characterize the $5 \bot 5 \bot 5$ intersection, while for $D4$ $D4$ $NS5$, only $3$ moduli are left, since the motion along $x_{2}$ is frozen.}.

\begin{figure}[htbp]
\begin{center}
\includegraphics[width=10cm, height=4cm]{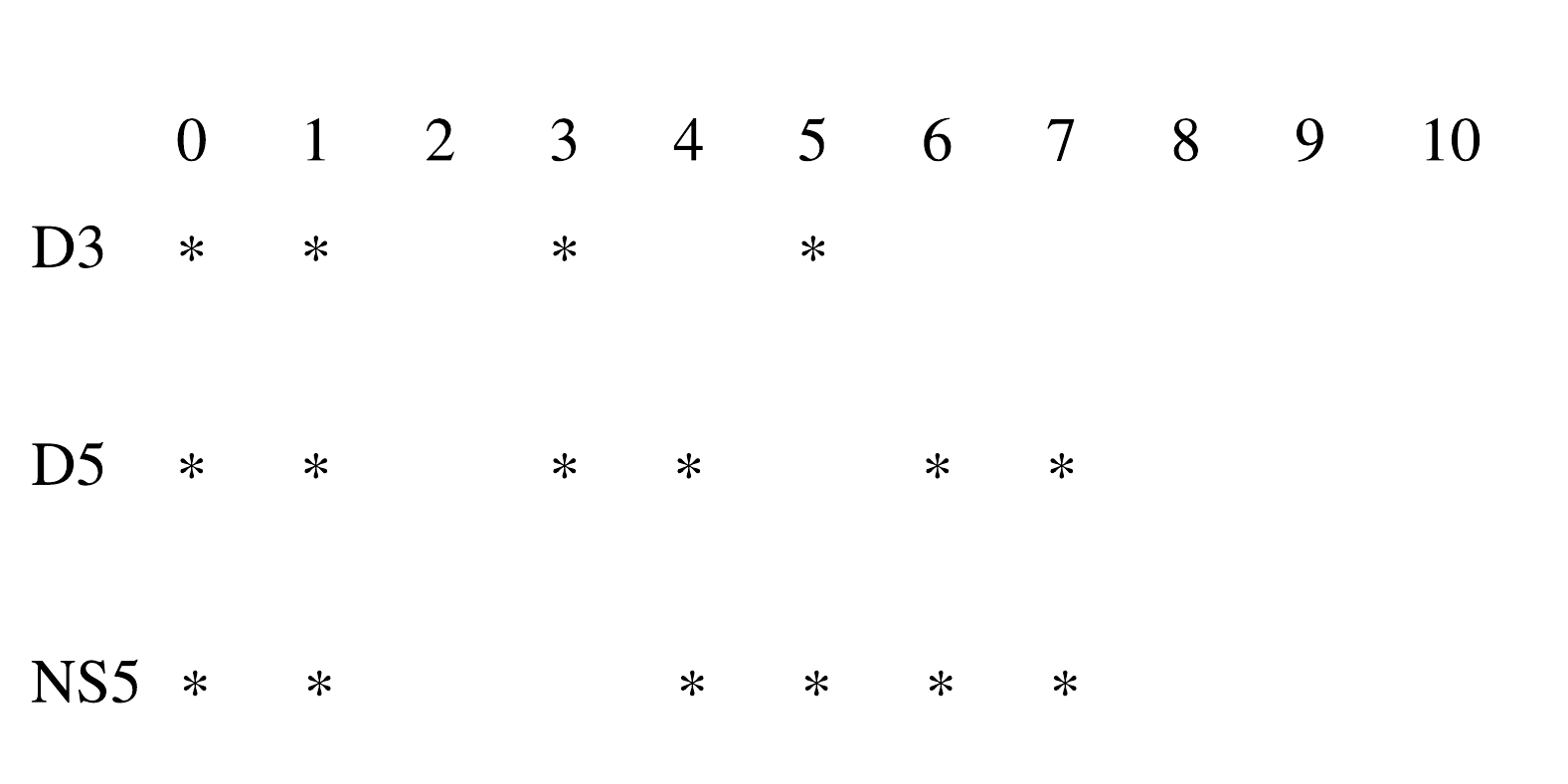}
\caption{The $D3$ $D5$ $NS5$ configuration}
\label{fig:2}
\end{center}
\end{figure}
Then compactify on $x_{4}$ with the radius $R_{4}$ and do a T-duality transformation, we get $D3$ $D5$ $NS5$ (see Fig.3). The state carrying $[i,j,k]$ index is the 3-string junction with $(p,q)$, $(p,0)$, $(0,q)$ strings ending on $D3$ $D5$ $NS5$, which will become massless when $v^{I}_{i}=v^{I}_{j}=v^{I}_{k}$. This is the scenario discussed in \cite{opk}. The 3-string junction is the point-like particle in $x_{0} \times x_{1}$, so they may give the field $f^{ijk}(x_{0},x_{1})$ localized at the intersection. In M theory with $x_{2} \times x_{4}$ compactified to $T^{2}$, the 3-string junction is lifted to a $M2$ embedded along a holomorphic curve in $x_{2} \times x_{4}\times x_{8}\times x_{9}\times x_{10} $, ending on the three $M5$'s along $(pR_{2},qR_{4})$, $p R_{2}$, $qR_{4}$ \cite{qwert}. Still, the problem is that when the three $M5$ branes intersect, the 3-string junction is at the threshold and may decay into the component strings. If they do decay, then at the triple intersection, there will be no BPS state related with all three branes.

\begin{figure}[htbp]
\begin{center}
\includegraphics[width=10cm, height=4cm]{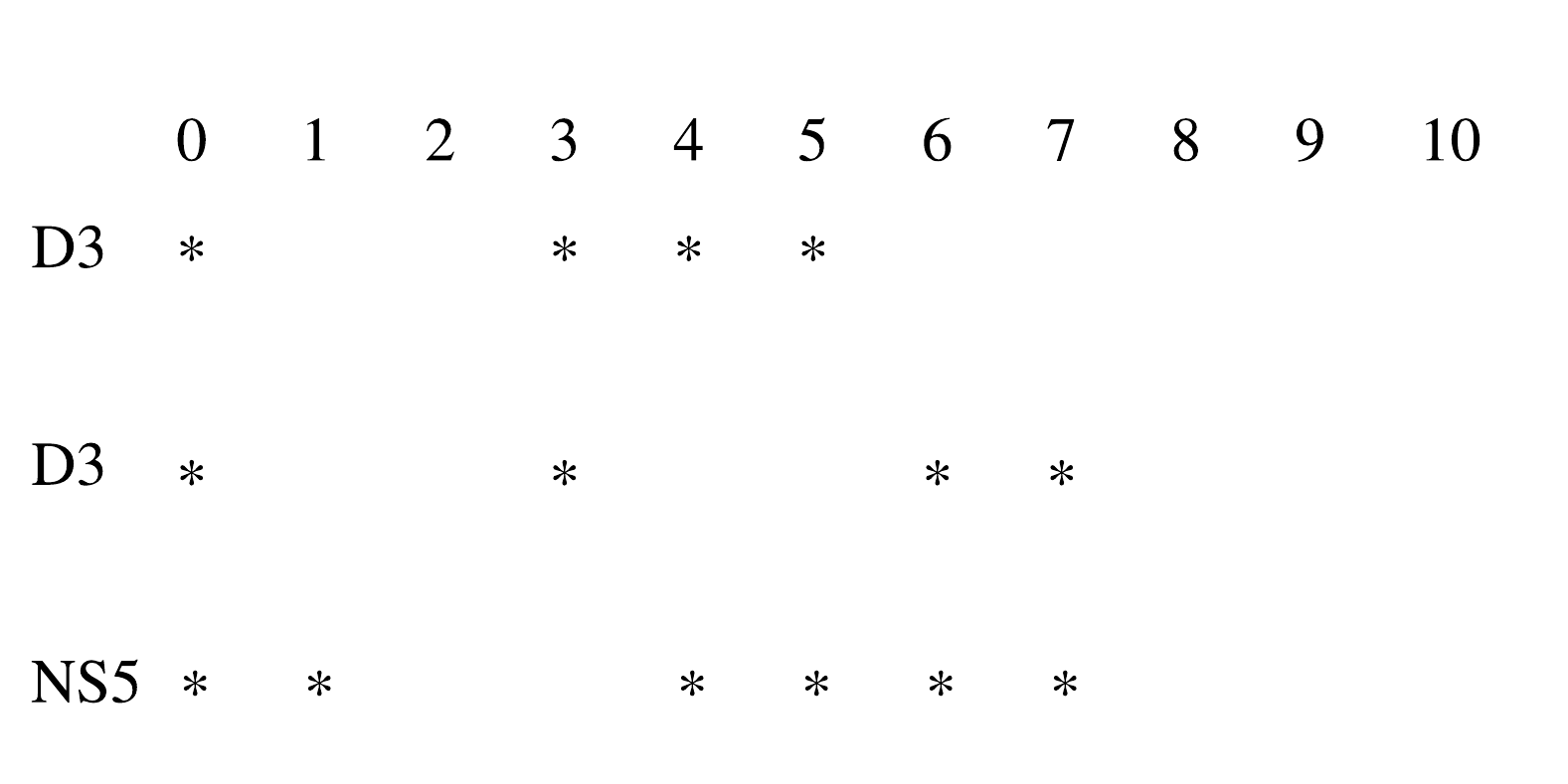}
\caption{The $D3$ $D3$ $NS5$ configuration}
\label{fig:3}
\end{center}
\end{figure}
The other possibility is to compactify on $x_{1}$ with the radius $R_{1}$ and also do a T-duality transformation. We get $D3$ $D3$ $NS5$ (see Fig.4). In $x_{8}\times x_{9} \times x_{10}$, no string junction can be formed. We may consider the 3-string junction in, for example, $x_{1}\times x_{8}$ plane. The KK mode along $x_{1}$ cannot be dropped. Actually, in the T-dual picture, $x_{1}$ is a circle with the radius $\frac{1}{T_{F1}R_{1}}$, so $D3$ and $D3$ will be separated with the distance $\frac{m}{T_{F1}R_{1}}$ in the covering space.

\begin{figure}[htbp]
\begin{center}
\includegraphics[width=10cm, height=4cm]{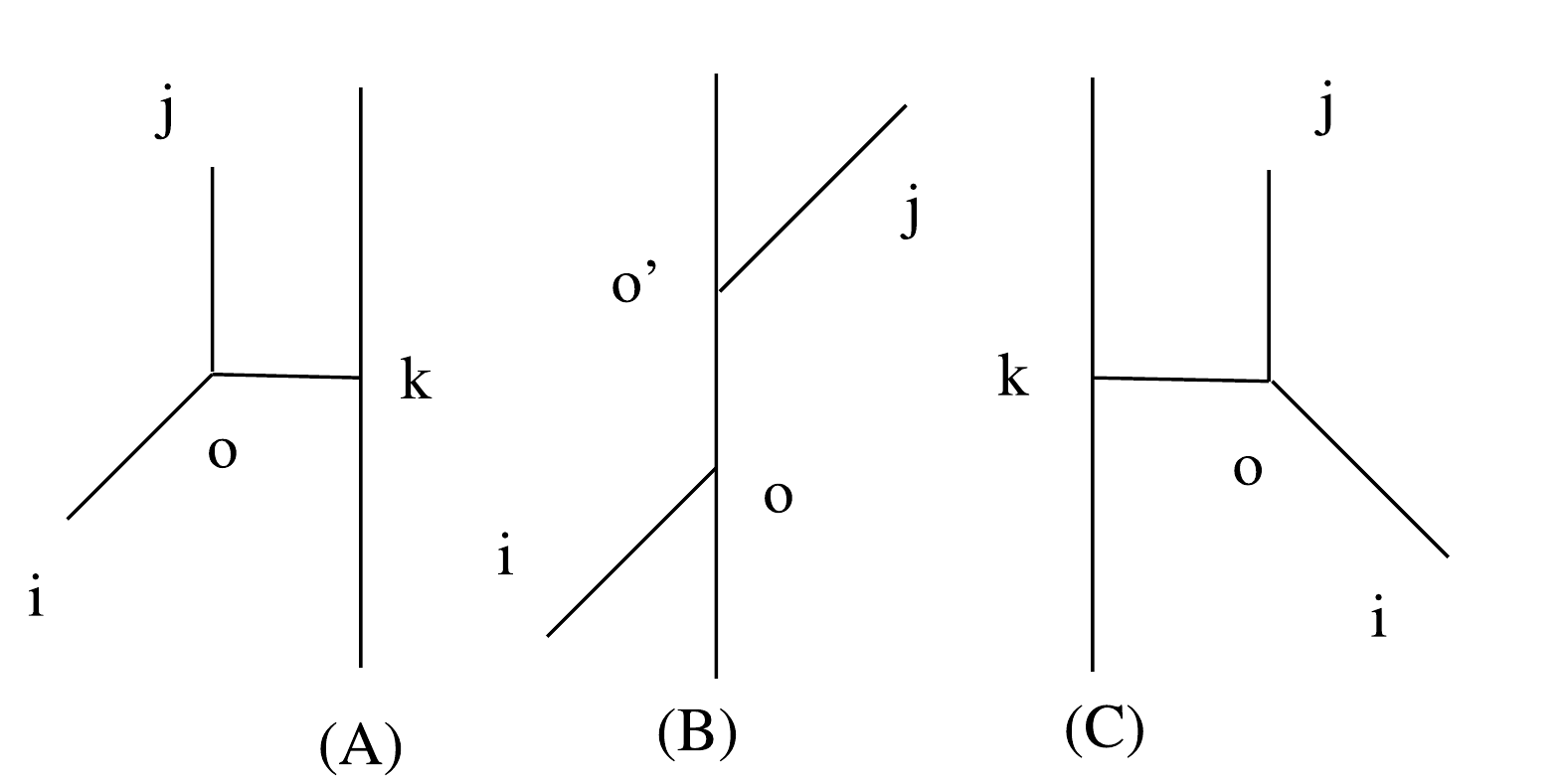}
\caption{The 3-string junction in $18$ plane}
\label{fig:4}
\end{center}
\end{figure}
In $x_{1}\times x_{8}$ plane, $NS5$ is a straight line locating in $v^{8}_{k}$, while the $i$ $j$ $D3$'s appear as two points with coordinates $(v^{1}_{i},v^{8}_{i})$, $(v^{1}_{j},v^{8}_{j})$, $v^{1}_{ij} = \frac{m}{T_{F1}R_{1}}$. The simplest 3-string junctions are given in Fig.5. In Figure 5 (A) and (C), the $io$ $oj$ $ok$ strings carry the charge $(p,q)$ $(p,0)$ $(0,q)$, $\tan \angle ioj = -\frac{qR_{1}}{pR_{2}}$. In Figure 5 (B), the $io$ $oo'$ $o'j$ strings carry the charge $(p,q)$ $(p,0)$ $(p,q)$, $\tan \angle ioo' = \tan \angle oo'j= -\frac{qR_{1}}{pR_{2}}$. Actually, there are also $(0,q)$ $oa$ string and the $(0,q)$ $bo'$ string ending on $NS5$ with the zero length. The string junctions like this always exist. The mass of the string junctions in (A) and (C) is $q T_{D1}|v^{8}_{ik}|+pT_{F1}|v^{1}_{ij}|$. The mass of the string junctions in (B) is $q T_{D1}|v^{8}_{ij}|+pT_{F1}|v^{1}_{ij}|$. In the T-dual picture, Figure 5 (A) corresponds to $q$ $[i,k]$ monopole strings with tension $T_{M2}|v^{8}_{ik}|$ wrapping $x_{1}$ carrying the $[i,j]$ longitudinal momentum $P_{1} = pm/R_{1}$. The situation is similar for Figure 5 (C). Figure 5 (B) corresponds to $q$ $[i,j]$ monopole strings with tension $T_{M2}|v^{8}_{ij}|$ wrapping $x_{1}$ carrying the $[i,j]$ longitudinal momentum $P_{1} = pm/R_{1}$.

When $v^{8}_{i}=v^{8}_{j}=v^{8}_{k}$, Figure 5 (A) and (C) reduce to the $[i,k]$ tensionless monopole strings wrapping $x_{1}$ carrying the $[i,j]$ longitudinal momentum $P_{1} = m/R_{1}$, which is offered by the potentially existing $[i,j]$ massless string. In Figure 5 (B), with $oa$ $oo'$ or $oo'$ $bo'$ kept, the state becomes the $[i,k]$ or $[k,j]$ tensionless monopole string wrapping $x_{1}$ carrying the $[i,j]$ longitudinal momentum $P_{1} = m/R_{1}$ offered by the $[i,j]$ massless string. Monopole string wrapping $x_{1}$ carrying $P_{1}$ momentum corresponds to the KK mode of $f^{ijk}(x_{0},x_{1})$ along $x_{1}$. Again, at the intersection, the state may decay into the monopole string and the string, and then there will be no BPS state relevant to all three branes.

At the $5 \bot 5 \bot 5$ intersection of three $NS5_{A}$'s or $NS5_{B}$'s or $D5$'s, there are type IIA strings, or type IIB strings, or D-strings living at the intersection. The F-string or D-string has the $4d$ transverse monition thus may produce the $c=6$ central charge. The oscillation of the F-string or D-string gives the $P_{1}$ momentum. The problem is that neither F-string nor D-string carries charge, so it is difficult to explain their relation with the three intersecting branes.

\section{Discussion}

In this paper, we considered the momentum modes of the $M5$ branes on a plane, which are the transverse momentum of the selfdual strings parallel to that plane. Different from the D branes, on which, the momentum modes are carried by the same kind of point-like excitations, here, the unparallel momentum modes are carried by selfdual strings with the different orientations. Selfdual strings with the same orientation gives a $5d$ SYM theory with the field configurations taking the zero Pontryagin number. The original $6d$ $(2,0)$ tensor multiplet field is then decomposed into a series of $\theta$-parameterized $5d$ $U(N)$ SYM fields, among which, fields labeled by the same $\theta$ have the standard SYM-type interaction. Fields labeled by different $\theta$ are associated with the selfdual strings with the different orientations. As a result, the $[i,j]+[j,k]\rightarrow [i,k]$ relation is not valid and the coupling cannot be realized as the standard matrix multiplication.

Since the bound state of the $[i,j]$ $\theta_{1}$ selfdual string and the $[j,k]$ $\theta_{2}$ selfdual string is not some $[i,k]$ selfdual string but the 3-string junction, we may also include the string junction into the theory. Each 3-string junction is characterized by $(\theta_{1},\theta_{2},\theta_{3})$, forming the tri-fundamental or anti-tri-fundamental representation of $U(N)$, and may couple with the $\theta_{1}$ $\theta_{2}$ $\theta_{3}$ selfdual strings in adjoint representation of $U(N)$. $[i,l]+[l,j,k]\rightarrow [i,j,k]$, $[j,m]+[i,m,k]\rightarrow [i,j,k]$, $[k,n]+[i,j,n]\rightarrow [i,j,k]$.

The quantization of the 3-string junction will give the higher-spin multiplet, for which the simplest one is the $(2,1)$ multiplet with the highest spin $3/2$. It is unclear whether the introducing of the 3-string junction will solve the problem or bring more problems, since at the beginning, we only want to get a theory for the $(2,0)$ tensor multiplet. The incorporation of the $5d$ massive $(2,1)$ multiplet into the $6d$ $(2,0)$ theory compactified on $S^{1}$ was also discussed in \cite{1111}, where it was suggested that the algebraic structure of the $6d$ $(2,0)$ theory may have a fermionic symmetry in addition to the self-dual tensor gauge symmetry.

Each 3-string junction carries three indices, so they may offer the $N^{3}$ degrees of freedom on $N$ $M5$ branes. However, the existing of the 3-string junction is severely restricted by the marginal stability curve, outside of which, the string junction may decay into the strings. For the given vacuum expectation values of the scalar fields on $M5$, the momentum of the string junction on that plane cannot be arbitrary. Especially, on coincident $M5$ branes, the 3-string junctions are at best at the marginal stability curve, so it is quite likely that they may decay.

Maybe it is easier consider the problem in the dual $D3$ picture. For $D3$ with the transverse dimension $x'^{45}$ compactified, the winding mode of the $(p,q)$ strings is dual to the $(n/R_{4},m/R_{5})$ momentum mode on $M5$. For the give $p$ and $q$, open $(p,q)$ strings with the arbitrary winding numbers have the SYM interaction. Then the questions are whether the open $(p,q)$ $(r,s)$ strings can interact or not, if can, in which way, what is the situation when $D3$ branes are coincident.

Among all selfdual strings, only those parallel to a given plane are taken as the perturbative degrees of freedom; nevertheless, different planes give the dual theories. One may compare the $6d$ theory with the $5d$ and $4d$ theories coming from the reductions on $x_{5}$ and $x_{4} \times x_{5}$. Obviously, selfdual string extending along $x_{5}$ is the only candidate to define the $5d$ theory. However, for $4d$ theory, any selfdual string parallel to the $45$ plane, carrying zero transverse momentum along it can act as the perturbative degrees of freedom. Only one is selected to give the $4d$ field, while the rest ones define the dual theories. Similarly, for $6d$ theory, selfdual strings parallel to a specific plane can be taken to give the $6d$ field, while the other planes give the dual versions. $M5$ on $S_{1}\times S_{2}\times S_{3}\times S_{4}\times S_{5}$ is dual to $D3$ on $S_{i}\times S_{j}\times S_{k}$ with a transverse $S'^{lm}$, where $\left\{1,2,3,4,5 \right\}=\left\{ i,j,k,l,m \right\}$. The $(p,q)$ string ending on $D3$ winding $S'^{lm}$ is dual to the selfdual string extending in $(qR_{l},pR_{m})$ direction, carrying transverse momentum in $x_{l} \times x_{m}$, localized in $x_{i} \times x_{j}\times x_{k}$. There are $10$ possible dual theories, corresponding to choosing the selfdual strings parallel to $10$ different $2d$ subspaces.

$M5$ on $T^{5}$ is $SL(5,Z)$ invariant. However, the $6d$ theory on $M5$ does not have the explicit $SL(5,Z)$ invariance. The $SL(5,Z)$ U-duality transformation, or the $SO(5)$ U-duality transformation in $R^{5}$, is not a simple differorphism transformation but is also accompanied by a reallocation of the perturbative and the nonpertubative degrees of freedom. The U-dual $6d$ theories are equivalent, so the $SL(5,Z)$ transformation is just like a change of the gauge. This is similar with the $D3$. Although $D3$ is S-duality invariant, the $4d$ theory on $D3$ does not have the $SL(2,Z)$ invariance. The nonpertubative $SL(2,Z)$ transformation gives the equivalent $4d$ theories.

\bigskip
\bigskip

{\bf Acknowledgments:}
The work is supported by the Mitchell-Heep Chair in High Energy Physics and by the DOE grant DE-FG03-95-Er-40917.

\bibliographystyle{plain}

\end{document}